\documentclass[aps,pra,twocolumn,showpacs,superscriptaddress]{revtex4-1} 

\usepackage{graphicx}
\usepackage[colorlinks,citecolor=blue,urlcolor=blue,linkcolor=blue]{hyperref}
\usepackage{siunitx}	
\usepackage{amsmath}
\usepackage{amsthm}
\usepackage{amssymb}
\usepackage{amsthm}
\usepackage{braket}
\usepackage{booktabs}
\usepackage{footmisc}
\usepackage{mathrsfs}
\usepackage{cleveref}
\usepackage{dsfont}
\usepackage{subfigure}
\usepackage{multirow}
\usepackage{enumerate}
\usepackage{lineno}
\usepackage{xcolor}
\definecolor{darkgreen}{rgb}{0, 0.5, 0}
\DeclareFontFamily{OT1}{pzc}{}
\DeclareFontShape{OT1}{pzc}{m}{it}{<-> s * [1.10] pzcmi7t}{}
\DeclareMathAlphabet{\mathpzc}{OT1}{pzc}{m}{it}

\newcommand{\unit}{\mathds{1}}

\newcommand{\sgn}{\mathrm{sgn}}

\newcommand{\tr}[1]{\mathrm{Tr}\left(#1\right)}
\newcommand{\trp}[2]{\mathrm{Tr}_\mathrm{#1}\left(#2\right)}
\newcommand{\abs}[1]{\left| #1 \right|} 

\newcommand{\bracket}[2]{\ensuremath{\langle#1 \vphantom{#2}| #2\vphantom{#1}\rangle}}

\newcommand{\ketbra}[2]{\ensuremath{|#1 \vphantom{#2}\rangle \langle #2\vphantom{#1}|}}

\begin{document}
\title{Characterizing four-body indistinguishability via symmetries}

\author{Alexander M. Minke}
\affiliation{Physikalisches Institut, Albert-Ludwigs-Universit{\"a}t Freiburg, Hermann-Herder-Str. 3, 79104 Freiburg, Germany}

\author{Andreas Buchleitner}
\affiliation{Physikalisches Institut, Albert-Ludwigs-Universit{\"a}t Freiburg, Hermann-Herder-Str. 3, 79104 Freiburg, Germany}
\affiliation{EUCOR Centre for Quantum Science and Quantum Computing, Albert-Ludwigs-Universität Freiburg, Hermann-Herder-Str. 3, 79104 Freiburg, Germany}

\author{Christoph Dittel}
\email{christoph.dittel@physik.uni-freiburg.de}
\affiliation{Physikalisches Institut, Albert-Ludwigs-Universit{\"a}t Freiburg, Hermann-Herder-Str. 3, 79104 Freiburg, Germany}
\affiliation{EUCOR Centre for Quantum Science and Quantum Computing, Albert-Ludwigs-Universität Freiburg, Hermann-Herder-Str. 3, 79104 Freiburg, Germany}

\date{\today}

\begin{abstract}
We show how to characterize the indistinguishability of up to four identical, bosonic or fermionic particles, which are rendered partially distinguishable through their internal degrees of freedom prepared in mixed states. This is accomplished via their counting statistics when subjected to a highly symmetric unitary acting upon their external (i.e., dynamical) degrees of freedom. For pure internal states, we further extract information on the particles' collective phases, which ultimately allows for an experimental reconstruction of the full many-particle density operator up to complex conjugation. 
\end{abstract}

\maketitle
\section{Introduction}
The indistinguishability of identical particles with respect to their internal degrees of freedom is \emph{the} essential ingredient for many-body interference in their external (i.e., dynamical) degrees of freedom \cite{Hong-MS-1987,Shih-NT-1988,Mayer-CS-2011}. This purely quantum mechanical interference effect can be exploited for a variety of applications, ranging from quantum simulations with ultracold atoms \cite{Bloch-QS-2012,Kaufman-TW-2014,Preiss-SC-2015,Islam-ME-2015,Gross-QS-2017,Kaufman-TH-2018,Preiss-HC-2019,Bergschneider-EC-2019} to photonic quantum information processing \cite{OBrien-OQ-2007,OBrien-PQ-2009,Aaronson-CC-2013,Carolan-UL-2015,Flamini-PQ-2018,Slussarenko-PQ-2019}. However, as a natural consequence of experimental limitations, real-world implementations commonly suffer from imperfect state preparation causing the particles to appear at least to some degree distinguishable \cite{Tichy-SP-2015,Shchesnovich-PI-2015,Tillmann-GM-2015,Walschaers-FM-2016,Khalid-PS-2018,Dittel-WP-2018}. Since the particles' external (i.e., dynamical) degrees of freedom are usually well controlled, there is considerable interest in utilizing interference patterns in the external degrees of freedom to characterize the particles' indistinguishability, without addressing those internal degrees of freedom which render them distinguishable. While for two particles such characterization is commonly achieved by measuring the coincidence rate in the Hong-Ou-Mandel experiment \cite{Hong-MS-1987,Shih-NT-1988}, the indistinguishability of more than two identical particles can give rise to genuine many-body properties \cite{Shchesnovich-CP-2018} that cannot be detected via two-particle Hong-Ou-Mandel interference between all pairs of particles.

In the literature several methods are known how to extract some -- but in general not all -- information on the particles' indistinguishability from characteristic signatures in the output statistics. This includes correlation measurements after random unitary transformations \cite{Walschaers-SB-2016,Walschaers-FM-2016,Giordani-ES-2018,Walschaers-SM-2020,Brunner-CD-2021}, imperfectly suppressed output events in generalizations of the Hong-Ou-Mandel experiment \cite{Tichy-ZT-2010,Tichy-MP-2012,Crespi-SL-2015,Weimann-IQ-2016,Dittel-MB-2017,Viggianiello-EG-2018,Tschernig-MD-2018,Dittel-TD1-2018,Dittel-TD2-2018,Ehrhardt-EC-2020,Muenzberg-SI-2021}, witnessing genuine many-body indistinguishability \cite{Brod-WG-2019,Giordani-EQ-2020}, or bounds on the interference visibility which are controlled by the particles' indistinguishability \cite{Shchesnovich-TB-2015,Stanisic-DD-2018,Dittel-WP-2018,Viggianiello-OP-2018}. However, to date it remains unknown how to exhaustively characterize many-body indistinguishability including the extraction of the particles' collective phases \cite{Shchesnovich-CP-2018} which non-trivially affect interference patterns in the dynamical degrees of freedom \cite{Menssen-DM-2017,Jones-ID-2020}.

Here we show how to fully characterize the indistinguishability of four identical bosonic or fermionic particles with arbitrary, independently prepared mixed states in the internal degrees of freedom, by analyzing the counting statistics after interference in the particles external degrees of freedom mediated by a non-interacting unitary transformation related to the hypercube symmetry. In particular, we characterize the degree of indistinguishability between all subsets of particles, and analyse which contributions stem from two-, three-, and four-particle properties. For pure internal states, we show how to extract the particles' collective phases, which express genuine many-body properties. These phases can be identified up to a global sign -- a limitation stemming from the symmetries of the unitary -- and allow for the reconstruction of the full many-body density operator up to complex conjugation. Altogether, this defines a comprehensive protocol for the detailed characterization and certification of many-body indistinguishability. It is applicable to a variety of experimental platforms, such as ultracold atoms in optical lattices \cite{Bloch-QS-2012,Kaufman-TW-2014,Preiss-SC-2015,Islam-ME-2015,Gross-QS-2017,Kaufman-TH-2018,Preiss-HC-2019,Bergschneider-EC-2019} or photons in linear optical networks \cite{OBrien-OQ-2007,OBrien-PQ-2009,Aaronson-CC-2013,Carolan-UL-2015,Flamini-PQ-2018,Slussarenko-PQ-2019}, and can readily be implemented in state-of-the-art photonics using multiport interferometers and photon-number-resolving detectors.

The manuscript is structured as follows: In Sec.~\ref{sec:ParDist} we start with the description of four partially distinguishable particles with uncorrelated mixed states in the internal degrees of freedom, and discuss in detail how to quantify the particles' indistinguishability. The unitary transformation and the subsequent detection process are investigated in Sec.~\ref{sec:Trans}, where we identify different classes of output events related to the hypercube symmetries, and derive expressions for all event probabilities in terms of indistinguishability quantifiers. In Sec.~\ref{sec:Chraracterizing} we then show how to extract the particles' indistinguishability from these output statistics, and, in Sec.~\ref{sec:Pure-internal-states}, we ultimately discuss the reconstruction of the full many-body density operator in the case of pure internal states. We conclude in Sec.~\ref{sec:Con}. For the sake of readability, all extensive proofs are deferred to the Appendices.


\section{Four-particle indistinguishability} \label{sec:ParDist}
We set the scene by introducing our formalism for the description of many identical particles prepared in mixed states in their internal degrees of freedom. The degree of (in-)distinguishability of the particles is fully controlled by their internal, their interference is probed in the external degrees of freedom. We discuss in detail how to quantify the particles' indistinguishability via specific characteristics of the many-body density operator. This establishes the basis for our inference of a four-particle state's indistinguishability properties as elaborated in the subsequent sections.

\subsection{Four partially distinguishable particles}\label{sec:FourPaSt}
Let us start with four identical bosons or fermions whose degrees of freedom are subdivided into \emph{external} ($\mathrm{E}$) and \emph{internal} ($\mathrm{I}$) degrees of freedom. While the former are subject to unitary transformations (therefore also called \emph{dynamical} degrees of freedom) described further down, we suppose that the latter are not acted upon -- neither by some transformation, nor by measurement. Thus, the particles' internal degrees of freedom solely serve to render them (partially) distinguishable. 

We consider four modes, which support the particles' external states, each initially occupied by exactly one particle, as described by the \emph{input mode occupation list} $\vec{R}=(1,1,1,1)$, with $R_j$ the number of particles in mode $j$. By labeling the particles from $1$ to $4$, their distribution among the modes can also be expressed via the \emph{input mode assignment list} $\vec{E}=(1,2,3,4)$, with $E_\alpha=\alpha$ the mode occupied by the $\alpha$th particle [see Fig.~\ref{fig:HC}(a) below]. Accordingly, the $\alpha$th factor in the tensor product 
\begin{align}\label{eq:E}
\ket{\vec{E}}=\ket{1,2,3,4}=\ket{1} \otimes  \ket{2}\otimes \ket{3} \otimes \ket{4}
\end{align}
corresponds to the external state of (i.e., the mode occupied by) the $\alpha$th particle, with $\bracket{\vec{E}}{\vec{E}'}=\prod_\alpha\delta_{E_\alpha,E_\alpha'}$, and $\delta_{j,k}$ the Kronecker delta. On the other hand, we make no restrictions on the internal state of each particle and merely assume the particles to be uncorrelated in their internal degrees of freedom. Consequently, the internal many-body density operator reads
\begin{align}\label{eq:rho}
\rho=\rho_1\otimes \rho_2\otimes \rho_3\otimes \rho_4,
\end{align}
with $\rho_\alpha$ the internal density operator of the $\alpha$th particle.

Let us now combine the description of the particles' external and internal degrees of freedom, and consider their tensor product $\ketbra{\vec{E}}{\vec{E}} \otimes \rho$. Given that we consider identical bosons (fermions), their common state must be (anti)symmetric with respect to the exchange of any two particles. This can be achieved by (anti)symmetrizing $\ketbra{\vec{E}}{\vec{E}} \otimes \rho$ with respect to the particles' labels -- given by the order in the tensor products in Eqs.~\eqref{eq:E} and~\eqref{eq:rho}. To this end we introduce the permutation operator $\Pi_\pi$ corresponding to $\pi\in\mathrm{S}_4$, with $\mathrm{S}_4$ the symmetric group of four elements, and permutations $\pi$ representing different orderings of the particles, which we call \emph{particle labellings}. The permutation operator corresponding to $\pi\in\mathrm{S}_4$ permutes the particles, it acts on the external states as
\begin{align*}
\Pi_\pi\ket{\vec{E}}=\ket{\pi(1),\pi(2),\pi(3),\pi(4)}=\ket{\vec{E}_\pi},
\end{align*}
and similarly on the internal states. Accordingly, with the (anti)symmetrization operator $\Pi_\mathrm{B(F)}=1/\sqrt{24} \sum_{\pi\in\mathrm{S}_4} (-1)^\pi_\mathrm{B(F)} \Pi_\pi \otimes \Pi_\pi$, the (anti)symmetrization of $\ketbra{\vec{E}}{\vec{E}} \otimes \rho$ yields
\begin{align}
\rho_\mathrm{EI}&=\Pi_\mathrm{B(F)} \left( \ketbra{\vec{E}}{\vec{E}} \otimes \rho\right) \Pi_\mathrm{B(F)}^\dagger \nonumber \\
&=\frac{1}{24} \sum_{\pi,\pi' \in \mathrm{S}_4} (-1)_\mathrm{B(F)}^{\pi\pi'} \ketbra{\vec{E}_\pi}{\vec{E}_{\pi'}} \otimes \Pi_\pi \rho \Pi_{\pi'}^\dagger,\label{eq:rhoIE}
\end{align}
where $(-1)_\mathrm{B}^{\pi\pi'}=1$ for bosons (B), $(-1)_\mathrm{F}^{\pi\pi'}=\sgn(\pi\pi')$ for fermions (F), and $\pi\pi'$ is the composition of $\pi$ with $\pi'$. Note that further below we also use the shorthands $\pm$ and $\mp$, with the upper (lower) sign applying to bosons (fermions). By Eq.~\eqref{eq:rhoIE} we now obtained an (anti)symmetric state, describing four identical bosons (fermions).

Since the dynamics and the detection of the particles [discussed in Sec.~\ref{sec:Trans} below] only acts upon their external degrees of freedom, we can consider the particles' external state obtained by tracing over the particles' internal degrees of freedom in~\eqref{eq:rhoIE}, 
\begin{align}\label{eq:rhoE}
\rho_\mathrm{E}=\trp{I}{\rho_\mathrm{EI}}=\sum_{\pi,\pi' \in \mathrm{S}_4} [\rho_\mathrm{E}]_{\pi,\pi'} \ketbra{\vec{E}_\pi}{\vec{E}_{\pi'}},
\end{align}
with
\begin{align}\label{eq:rhoElementInt}
[\rho_\mathrm{E}]_{\pi,\pi'}=(-1)_\mathrm{B(F)}^{\pi\pi'} \frac{1}{24} \tr{\Pi_\pi \rho \Pi_{\pi'}^\dagger }.
\end{align}
Using $\Pi_{\pi'}^\dagger \Pi_\pi=\Pi_\kappa$ for $\kappa=\pi(\pi')^{-1}$, and the cyclic property of the trace, the matrix elements~\eqref{eq:rhoElementInt} can be written as
\begin{align}\label{eq:rhoEPiKappa}
[\rho_\mathrm{E}]_{\pi,\pi'}&=[\rho_\mathrm{E}]_{\kappa,\epsilon},
\end{align}
with $\epsilon$ the identity. 

Note that, by Eq.~\eqref{eq:rhoElementInt}, for fully distinguishable $\mathrm{(D)}$ particles, for which $\rho_\alpha$ and $\rho_\beta$ have orthogonal support for all $\alpha \neq \beta$, we have $[\rho_\mathrm{E}^\mathrm{D}]_{\pi,\pi'}= \delta_{\pi,\pi'}/24$, and, accordingly, a diagonal reduced external state
\begin{align}\label{eq:rhoED}
\rho_\mathrm{E}^\mathrm{D}=\frac{1}{24}\sum_{\pi\in\mathrm{S}_4} \ket{\vec{E}_\pi}\bra{\vec{E}_\pi}.
\end{align}
On the other hand, in the case of perfectly indistinguishable bosons (fermions), the particles' internal states $\rho_\alpha$ are pure and equal for all $\alpha\in\{1,2,3,4\}$, yielding $[\rho_\mathrm{E}^\mathrm{B(F)}]_{\pi,\pi'}=(-1)^{\pi\pi'}_\mathrm{B(F)}/24$. Hence, we obtain a pure state $\rho_\mathrm{E}^\mathrm{B(F)}=\ket{\psi_\mathrm{B(F)}}\bra{\psi_\mathrm{B(F)}}$, with \cite{Dittel-WP-2018,Dittel-AI-2019}
\begin{align}\label{eq:Indpurestate}
\ket{\psi_\mathrm{B(F)}}=\frac{1}{\sqrt{24}}\sum_{\pi\in\mathrm{S}_4} (-1)^{\pi}_\mathrm{B(F)} \ket{\vec{E}_\pi}.
\end{align}

\subsection{Quantifying indistinguishability}\label{sec:quant}
In order to see how the particles' indistinguishability can be quantified via characteristics of the reduced external state $\rho_\mathrm{E}$, let us inspect its matrix elements~\eqref{eq:rhoElementInt} in more detail. First of all, the permutation of particles via the operators $\Pi_\pi$ in Eq.~\eqref{eq:rhoElementInt} already reveals that the off-diagonal elements of $\rho_\mathrm{E}$ are governed by overlaps of internal states, hence the coherence of the many-particle state~\eqref{eq:rhoE} is controlled by the particles' indistinguishability. In particular, for indistinguishable particles, $\rho_\mathrm{E}^\mathrm{B(F)}$ is maximally coherent, $[\rho_\mathrm{E}^\mathrm{B(F)}]_{\pi,\pi'}=(-1)_\mathrm{B(F)}^{\pi\pi'}/24$, and for fully distiguishable particles there is no coherence since all off-diagonal elements vanish, $[\rho_\mathrm{E}^\mathrm{D}]_{\pi,\pi'}=\delta_{\pi,\pi'}/24$ \cite{Dittel-WP-2018}. 

The connection between the elements $[\rho_\mathrm{E}]_{\pi,\pi'}$ and the particles' indistinguishability becomes clearer by considering the cycle structure of $\kappa\equiv \pi(\pi')^{-1}$ [see Eq.~\eqref{eq:rhoEPiKappa}]. Suppose that $\kappa$ has $c$ cycles, $\kappa=\kappa_1 \cdots \kappa_c$, with the $j$th cycle $\kappa_j=(\kappa_{j,1} \cdots \kappa_{j,l_j})$ being of length $l_j$, and $\kappa_{j,\alpha} \in \{1,\dots,4\}$. We characterize the cycle structure of $\kappa$ by the lengths of its cycles, $L=(l_1,\dots, l_c)$. In number-theoretical terms, $L$ is called a partition \cite{Dummit-AA-2003} of $4$. For example $\kappa=(2)(1~4~3)$ has $c=2$ cycles, $\kappa_1=(2)$ and $\kappa_2=(1~4~3)$, with length $l_1=1$ and $l_2=3$, respectively, such that the cycle structure of $\kappa$ reads $L=(1,3)$. Note that in the following we drop cycles of length one and write, e.g., $\kappa=(2)(1~4~3)\equiv (1~4~3)$. 

With the help of the cycle structure of $\kappa=\pi(\pi')^{-1}$ we can write the matrix elements~\eqref{eq:rhoElementInt} of $\rho_\mathrm{E}$ as \cite{Shchesnovich-PI-2015}
\begin{align}
[\rho_\mathrm{E}]_{\pi,\pi'}&=(-1)_\mathrm{B(F)}^{\kappa}\frac{1}{24}\prod_{j=1}^c \tr{\rho_{\kappa_{j,1}} \cdots \rho_{  \kappa_{j,l_j}  }  },\label{eq:rhoElementTr}
\end{align}
where the argument of the trace is a \emph{matrix product} of internal single-particle density operators $\rho_\alpha$. In general, the trace $\tr{\rho_\alpha \rho_\beta \cdots \rho_\gamma}$ is complex, reading in polar representation
\begin{align}\label{eq:CollPhase}
\tr{\rho_\alpha \rho_\beta \cdots \rho_\gamma  }= T_{(\alpha \beta \cdots \gamma)} \ e^{\mathrm{i} \varphi_{(\alpha \beta \cdots \gamma)}},
\end{align}
with $T_{(\alpha \beta \cdots \gamma)}\in \mathbb{R}$, and $-\pi \leq \varphi_{(\alpha \beta \cdots \gamma)} < \pi$ the particles' \emph{collective phase} \cite{Shchesnovich-CP-2018}. Note that if only two density operators are involved, we have $\tr{\rho_\alpha \rho_\beta} \geq 0$, since $\rho_\alpha$ and $\rho_\beta$ are Hermitian and positive semi-definite \footnote{Since $\rho_\alpha$ and $\rho_\beta$ are Hermitian and positive semi-definite, we can use their eigendecomposition $\rho_\alpha =\sum_j a_j \ket{a_j}\bra{a_j}$ and $\rho_\beta =\sum_k b_k \ket{b_k}\bra{b_k}$, with $a_j,b_k \geq 0$, such that $\unexpanded{\tr{\rho_\alpha\rho_\beta}=\sum_{j,k} a_j b_k | \bracket{a_j}{b_k} |^2\geq 0}$. }, such that the collective phase vanishes, i.e., $\varphi_{(\alpha \beta)}=0$ for all $\alpha,\beta \in \{1,\dots,4\}$. Further note that, unlike a global phase, the collective phase can affect the particles' output statistics as it is inscribed in the state's off-diagonal elements and cannot be factorized out. This was recently observed for three \cite{Menssen-DM-2017} and four \cite{Jones-ID-2020} particles. 

The expression for the off-diagonal elements of $\rho_\mathrm{E}$ in Eq.~\eqref{eq:rhoElementTr} reveals that the cycle structure of $\kappa=\pi(\pi')^{-1}$ determines how these are related to the particles' indistinguishability. In particular, $[\rho_\mathrm{E}]_{\pi,\pi'}$ quantifies how well one can distinguish the particle labellings $\pi$ and $\pi'$, or, equivalently, by $\kappa=\pi(\pi')^{-1}$, the labellings $\kappa$ and $\epsilon$, with $\epsilon$ the identity permutation. The ability to discriminate the particles' labellings is then tantamount to distinguishing the particles via their internal states \cite{Dittel-WP-2018}. Recall that these artificial, non-physical particle labellings came along with Eqs.~\eqref{eq:E} and~\eqref{eq:rho}, and were subsequently `erased' via the (anti)symmetrization of the many-particle state~\eqref{eq:rhoIE}. 

Let us now inspect in more detail how the elements of $\rho_\mathrm{E}$ look like for the five different cycle structures $L$ occurring in $\mathrm{S}_4$:

\begin{enumerate}[(i)]
\item $L=(1,1,1,1)$: The only permutation $\kappa=\pi(\pi')^{-1}$ whose cycles are all of length one is the identity $\epsilon$. In this case we have equal labellings, $\pi=\pi'$, yielding $[\rho_\mathrm{E}]_{\pi,\pi}=1/24$. These are the diagonal elements of $\rho_\mathrm{E}$. 

\item $L=(1,1,2)$: If $\kappa$ is characterized by $L=(1,1,2)$, say $\kappa=(\alpha~\beta)$, we get $[\rho_\mathrm{E}]_{\pi,\pi'}=\pm \tr{\rho_\alpha\rho_\beta}/24$. This element characterizes the indistinguishability between particle labellings differing by an exchange of $\alpha$ and $\beta$. Recall that $\tr{\rho_\alpha\rho_\beta}$ is real and positive. 

\item $L=(1,3)$: The indistinguishability of labellings $\pi$ and $\pi'$ for which $\kappa=(\alpha~\beta~\gamma)$ is quantified by the element $[\rho_\mathrm{E}]_{\pi,\pi'}=\tr{\rho_\alpha\rho_\beta\rho_\gamma}/24\in \mathbb{C}$. The collective phase $\varphi_{(\alpha\beta\gamma)}$ of $\tr{\rho_\alpha\rho_\beta\rho_\gamma}$ is also called \cite{Menssen-DM-2017} \emph{triad phase}. Note that there is a second permutation $\kappa$ of length three, which involves the same three particles, $\kappa=(\alpha~\gamma~\beta)$. 

\item $L=(2,2)$: For $\kappa$ containing two cycles of length two, $\kappa=(\alpha~\beta)(\gamma~\delta)$, the corresponding element $[\rho_\mathrm{E}]_{\pi,\pi'}=\tr{\rho_\alpha\rho_\beta}\tr{\rho_\gamma\rho_\delta}/24$ quantifies the indistinguishability of labellings where particle $\alpha$ is exchanged with $\beta$, and  $\gamma$ with $\delta$.

\item $L=(4)$: Finally, in the case of a single cycle of length four, $\kappa=(\alpha~\beta~\gamma~\delta)$, the indistinguishability of the corresponding labellings is given by $[\rho_\mathrm{E}]_{\pi,\pi'}=\pm\tr{\rho_\alpha\rho_\beta \rho_\gamma\rho_\delta}/24$.  Again, there are several permutations $\kappa$ (precisely 6) containing all four particles in a single cycle. 
\end{enumerate}

For our characterization of four-particle indistinguishability it is useful to combine those matrix elements given in Eq.~\eqref{eq:rhoElementTr} which exhibit the same cycle structure $L$ of $\kappa$. To do so, we introduce the (non-normalized) indistinguishability measure
\begin{align}\label{eq:DL}
\mathcal{I}_{L}&=\sum_{\substack{\kappa \in \mathrm{S}_4 \\L(\kappa)=L }}\prod_{j=1}^c \tr{\rho_{\kappa_{j,1}} \cdots \rho_{  \kappa_{j,l_j}  }  },
\end{align}
where $L(\kappa)=L$ selects $\kappa$ with cycle structure $L$. Since the internal single-particle density operators $\rho_\alpha$ are Hermitian, we have that $\tr{\rho_\alpha \rho_\beta \cdots \rho_\gamma  }=\tr{\rho_\gamma \cdots \rho_\beta \rho_\alpha  }^*$ for inverse cycles $\kappa_j=(\alpha \beta \cdots \gamma)$ and $\kappa_j^{-1}=(\gamma \cdots \beta\alpha)$. Accordingly, by Eq.~\eqref{eq:CollPhase},
\begin{align}\label{eq:trcos}
\begin{split}
&\tr{\rho_\alpha \rho_\beta \cdots \rho_\gamma  }+\tr{\rho_\gamma \cdots \rho_\beta \rho_\alpha  }\\
=&\ 2 \ T_{(\alpha \beta \cdots \gamma)} \cos \varphi_{(\alpha \beta \cdots \gamma)} ,
\end{split}
\end{align}
and the indistinguishability measures $\mathcal{I}_{L}$ of all particle labellings with cycle structure $L$ [see Eq.~\eqref{eq:DL}] can be written as
\begin{widetext}
\begin{align}\label{eq:Dall}
\begin{split}
\mathcal{I}_{(1,1,2)}=&\ T_{(12)}+T_{(13)}+T_{(14)}+T_{(23)}+T_{(24)} +T_{(34)} , \\
\mathcal{I}_{(1,3)}=&\ 2\Big[ T_{(123)} \cos\varphi_{(123)}+ T_{(124)} \cos\varphi_{(124)}+ T_{(134)} \cos\varphi_{(134)}+ T_{(234)} \cos\varphi_{(234)}\Big],\\
\mathcal{I}_{(2,2)}=&\ T_{(12)}T_{(34)}+T_{(13)}T_{(24)}+T_{(14)}T_{(23)} , \\
\mathcal{I}_{(4)}=&\  2\Big[ T_{(1234)} \cos\varphi_{(1234)}+T_{(1243)} \cos\varphi_{(1243)} +T_{(1324)} \cos\varphi_{(1324)}\Big].
\end{split}
\end{align}
\end{widetext}

With the above tools at hand, let us now quantify the indistinguishability between all four particles and recall that {the reduced external state $\rho_\mathrm{E}$ is diagonal with no preferred symmetry in the case of fully distinguishable particles [see Eq.~\eqref{eq:rhoED}], and totally (anti)symmetric in the case of indistinguishable bosons (fermions) [see Eq~\eqref{eq:Indpurestate}]. Hence, as a measure for the particles' indistinguishability we utilize the expectation value of the projector $\Pi_\mathrm{B(F)}=1/4! \sum_{\pi \in \mathrm{S}_4} (-1)^\pi_\mathrm{B(F)}\Pi_\pi$ onto the  (anti)symmetric subspace,
\begin{align}
\braket{\Pi_\mathrm{B(F)}}_\mathrm{4p}&=\tr{\Pi_\mathrm{B(F)} \rho_\mathrm{E}} \label{eq:W4pdef}\\
&=\frac{1}{24}\sum_{\pi,\pi' \in \mathrm{S}_4 } (-1)^{\pi\pi'}_\mathrm{B(F)}\left[ \rho_\mathrm{E} \right]_{\pi,\pi'},\label{eq:W4pStart}
\end{align} 
with~\eqref{eq:W4pStart} obtained by plugging $\rho_\mathrm{E}$ from~\eqref{eq:rhoE} into~\eqref{eq:W4pdef}. It quantifies the particles' indistinguishability via the (anti)symmetry of $\rho_\mathrm{E}$, measured by the sum of all its matrix elements. Besides the off-diagonal elements, this also includes the diagonal elements, which are independent on the particles' indinstinguishability, and, thus, contribute a constant factor, such that $\braket{\Pi_\mathrm{B(F)}}_\mathrm{4p}=1/24$ for fully distinguishable particles, and $\braket{\Pi_\mathrm{B(F)}}_\mathrm{4p}=1$ for indistinguishable bosons (fermions). The sign in~\eqref{eq:W4pStart} compensates the sign in the matrix elements~\eqref{eq:rhoElementInt}. Hence, $\braket{\Pi_\mathrm{B(F)}}_\mathrm{4p}$ can be expressed equally for both bosons and fermions [see Eq.~\eqref{eq:W4p} below]. Note that in \cite{Brunner-CD-2021} a similar measure of the particles' indistinguishability is considered, differing from Eq.~\eqref{eq:W4pStart} by a normalisation constant and the sign factor. Further note that since $\ket{\psi_\mathrm{B(F)}}$ from Eq.~\eqref{eq:Indpurestate} is the eigenstate of $\rho_\mathrm{E}$ living in the (anti)symmetric subspace \cite{Dittel-WP-2018}, $\rho_\mathrm{E} \ket{\psi_\mathrm{B(F)}}=\lambda_\mathrm{B(F)} \ket{\psi_\mathrm{B(F)}}$, we can identify the expectation value of the projector onto the (anti)symmetric subspace of four particles with the corresponding eigenvalue, $\braket{\Pi_\mathrm{B(F)}}_\mathrm{4p}=\lambda_\mathrm{B(F)}$. Moreover, as shown in \cite{Dittel-WP-2018} [Eq.~(59) there], this eigenvalue is equivalent to the squared fidelity $F^2(\rho_\mathrm{E},\rho_\mathrm{E}^\mathrm{B(F)})$ between $\rho_\mathrm{E}$ and the state $\rho_\mathrm{E}^\mathrm{B(F)}$ corresponding to perfectly indistinguishable bosons (fermions) [see Eq.~\eqref{eq:Indpurestate}], i.e., $\lambda_\mathrm{B(F)}=F^2(\rho_\mathrm{E},\rho_\mathrm{E}^\mathrm{B(F)})=\braket{\Pi_\mathrm{B(F)}}_\mathrm{4p}$.

Let us now explicitly express the four-particle indistinguishability measure~\eqref{eq:W4pStart} in terms of the indistinguishability quantifiers from Eq.~\eqref{eq:Dall}. To this end, we rewrite Eq.~\eqref{eq:W4pStart} with the help of Eqs.~\eqref{eq:rhoElementTr},~\eqref{eq:DL}, and~\eqref{eq:Dall}, as
\begin{align}\label{eq:WMkappa}
\braket{\Pi_\mathrm{B(F)}}_\mathrm{4p}&=\sum_{\kappa \in \mathrm{S}_4} (-1)^\kappa_\mathrm{B(F)}\left[ \rho_\mathrm{E} \right]_{\kappa,\epsilon}\\
&=\frac{1}{24} \left[1+\mathcal{I}_{(1,1,2)}+\mathcal{I}_{(1,3)}+\mathcal{I}_{(2,2)}+\mathcal{I}_{(4)}\right].\label{eq:W4p}
\end{align}
Accordingly, $\braket{\Pi_\mathrm{B(F)}}_\mathrm{4p}$ involves elements~\eqref{eq:rhoElementTr} corresponding to all $4!=24$ labellings $\kappa$, i.e. it can be interpreted as fully quantifying the indistinguishability between all involved particle labellings.

To see which terms in~\eqref{eq:W4p} correspond to two- and three-particle properties, let us consider the expectation value $\braket{\Pi_\mathrm{B(F)}}_{N\mathrm{p}}=\tr{\Pi_\mathrm{B(F)} \rho_\mathrm{E}^{N\mathrm{p}}}$ of the reduced $N$-particle state $\rho_\mathrm{E}^{N\mathrm{p}}$ \cite{Brunner-CD-2021} obtained from $\rho_\mathrm{E}$ by tracing out all but $N\in\{2,3\}$ particles, $\rho_\mathrm{E}^{N\mathrm{p}}=\mathrm{Tr}_{4-N}(\rho_\mathrm{E})$. Using the projector $\Pi_\mathrm{B(F)}=1/N! \sum_{\pi \in \mathrm{S}_N} (-1)^\pi_\mathrm{B(F)}\Pi_\pi$ onto the (anti)symmetric subspace of $N$ particles, a straightforward calculation shown in Appendix~\ref{app:3p2p} reveals
\begin{align}\label{eq:Wa3p}
\braket{\Pi_\mathrm{B(F)}}_\mathrm{3p}&=\frac{1}{6} \left[1+ \frac{1}{2} \ \mathcal{I}_{(1,1,2)} + \frac{1}{4} \ \mathcal{I}_{(1,3)} \right],
\end{align}
and
\begin{align}\label{eq:Wa2p}
\braket{\Pi_\mathrm{B(F)}}_\mathrm{2p}&=\frac{1}{2}\left[1+\frac{1}{6} \ \mathcal{I}_{(1,1,2)} \right].
\end{align}
Similarly to $\braket{\Pi_\mathrm{B(F)}}_\mathrm{4p}$ from Eq.~\eqref{eq:W4pStart}, these measures satisfy  $1/N!\leq \braket{\Pi_\mathrm{B(F)}}_{N\mathrm{p}} \leq 1$, with the lower (upper) bound saturated in the case of fully distinguishable (indistinguishable) particles. In view of Eqs.~\eqref{eq:W4p},~\eqref{eq:Wa3p}, and~\eqref{eq:Wa2p} we see that the indistinguishability quantifiers $\mathcal{I}_{(1,1,2)}$ and $\mathcal{I}_{(1,3)}$ can be ascribed to two- and three-particle properties, respectively, while $\mathcal{I}_{(2,2)}$ and $\mathcal{I}_{(4)}$ only appear for the full four-particle state [see Eq.~\eqref{eq:W4p}].

While Eqs.~\eqref{eq:W4p}-\eqref{eq:Wa2p} refer to the properties of \emph{four} particles, we now restrict our analysis to a subset of $N\in\{2,3\}$ \emph{specific} particles $\{\alpha, \dots,\beta\}$. To this end, we consider their state $\rho_\mathrm{E}^{(\alpha, \dots,\beta)}$, which can be constructed similarly as the four-particle state $\rho_\mathrm{E}$ [see Sec.~\ref{sec:FourPaSt}], by merely considering particles $\{\alpha, \dots,\beta\}$. For the indistinguishability of particles $\{\alpha,\beta,\gamma\}$ and particles $\{\alpha,\beta\}$, we find
\begin{align}\label{eq:W3p}
\begin{split}
\braket{\Pi_\mathrm{B(F)}}_{(\alpha,\beta, \gamma)}= &\frac{1}{6}\big[ 1+ T_{(\alpha \beta)}+ T_{(\alpha \gamma)}+ T_{(\beta \gamma)}\\
&+2T_{(\alpha \beta \gamma)} \cos\varphi_{(\alpha \beta \gamma)} \big],
\end{split}
\end{align}
and 
\begin{align}
\braket{\Pi_\mathrm{B(F)}}_{(\alpha,\beta)}=\frac{1}{2}\left[1+ T_{(\alpha \beta)}\right],\label{eq:W2p}
\end{align}
which, similar to $\braket{\Pi_\mathrm{B(F)}}_\mathrm{4p}$ from Eq.~\eqref{eq:W4p}, quantifies the indistinguishability of particles $\{\alpha, \beta, \gamma\}$ and $\{\alpha ,\beta\}$ via the indistinguishability of their labellings, respectively.

\section{Transformation and output statistics} \label{sec:Trans}
In the previous section we demonstrated in detail how to quantify the indistinguishability of four particles by specific properties of the reduced external state $\rho_\mathrm{E}$. We now proceed with a discussion of the unitary transformation acting upon the particles, and of the subsequent detection protocol. These merely act on the particles' \emph{external} degrees of freedom, i.e. do not affect their internal states, through which they get (partially) distinguishable. However, since many-particle interference in the particles' dynamics depends on their degree of indistinguishability, the resulting output statistics encodes relevant information on this very property. How to extract this information without addressing the particles' internal states is discussed in Sec.~\ref{sec:Chraracterizing}.

\begin{figure}[t]
\centering
\includegraphics[width=\linewidth]{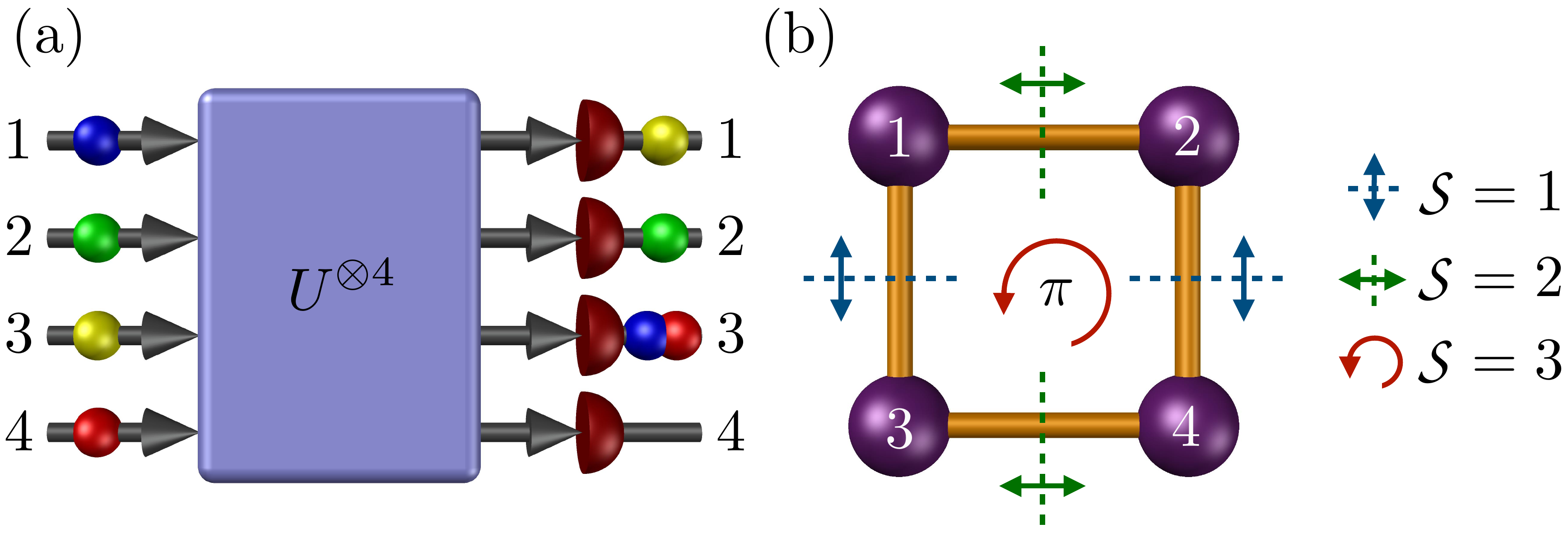}
\caption{Transformation under the two-dimensional hypercube unitary. (a) Four partially distinguishable particles (colored balls) in the input mode occupation $\vec{R}=(1,1,1,1)$ undergo a non-interacting unitary transformation defined by the single-particle hypercube unitary $U$ (light blue box). After the transformation, the output mode occupation by the particles is measured --  here illustrated for the output event with mode occupation list $\vec{S}=(1,1,2,0)$ and mode assignment list $\vec{F}=(1,2,3,3)$. (b) The coupling structure of the Hamiltonian (modes are represented by vertices, their couplings by edges) which generates the hypercube unitary exhibits three symmetries, $\mathpzc{S} = 1,2,$ and $3$, corresponding to the mode permutations $\sigma^{(1)}=(1~3)(2~4)$, $\sigma^{(2)}=(1~2)(3~4)$, and $\sigma^{(3)}=(1~4)(2~3)$, respectively.}
\label{fig:HC}
\end{figure}

\subsection{Transformation and measurement} \label{sec:trans}
As illustrated in Fig.~\ref{fig:HC}(a), we investigate a map $\rho_\mathrm{E}\rightarrow \mathcal{U}\rho_\mathrm{E}\mathcal{U}^\dagger$ acting upon the particles' external degrees of freedom. Here $\mathcal{U}=U^{\otimes 4}$ is a non-interacting unitary transformation governed by the two-dimensional single-particle hypercube unitary \cite{Dittel-MB-2017},
\begin{align}\label{eq:U}
U=e^{-\mathrm{i} Ht/\hbar}=\frac{1}{2} \begin{pmatrix} 1&\mathrm{i}\\\mathrm{i}&1\end{pmatrix}^{\otimes 2} =\frac{1}{2} \begin{pmatrix} 
1&\mathrm{i} &\mathrm{i} &-1   \\
\mathrm{i} &1 &-1 &\mathrm{i}   \\
\mathrm{i} &-1 &1 &\mathrm{i}    \\
-1&\mathrm{i} &\mathrm{i} &1   \end{pmatrix},
\end{align}
generated after time $t=-\pi/(4 c)$ by the single-particle Hamiltonian 
\begin{align*}
H=c \hbar\begin{pmatrix} 
0&1 &1 &0   \\
1 &0 &0 &1   \\
1 &0 &0 &1    \\
0&1 &1 &0   \end{pmatrix},
\end{align*}
which connects each mode $j$ with two neighbouring modes $k$ with constant coupling strength $c=H_{j,k}/\hbar$ [see Fig.~\ref{fig:HC}(b)].

A projective measurement of the particles in the output modes [see Fig.~\ref{fig:HC}(a)] gives rise to detection events with a specific particle distribution among the modes, described by the \emph{output mode occupation list} $\vec{S}$. The latter is defined in strict analogy to $\vec{R}$ [see Sec.~\ref{sec:FourPaSt}], i.e. $S_j$ counts the number of particles in mode $j$. Moreover, similar to $\vec{E}$ [see Sec.~\ref{sec:FourPaSt}], let us also define the \emph{output mode assignment list} $\vec{F}$, which lists the mode numbers occupied by the particles in ascending order [see Fig.~\ref{fig:HC}(a) for an example]. Together with the shorthand $S=24/\prod_j S_j!$, the measurement of the output mode occupations $\vec{S}$ is then realized by the projection operators
\begin{align*}
M_{\vec{S}}= \frac{S}{24} \sum_{\pi \in \mathrm{S}_4} \ketbra{\vec{F}_\pi}{\vec{F}_\pi},
\end{align*}
satisfying $M_{\vec{S}}^\dagger M_{\vec{S}}=M_{\vec{S}}$, and $\sum_{\vec{S}} M_{\vec{S}} = \unit$. Consequently, the probability for an output event $\vec{S}$ is given by \cite{Dittel-AI-2019}
\begin{align}
p_{\vec{S}}&= \tr{M_{\vec{S}} \ \mathcal{U} \rho_\mathrm{E} \mathcal{U}^\dagger } \nonumber \\
&=S \sum_{\pi, \pi' \in \mathrm{S}_4} [\rho_\mathrm{E}]_{\pi,\pi'} \prod_{\alpha=1}^4 U_{F_\alpha, \pi(\alpha)} U^*_{F_\alpha, \pi'(\alpha)},\label{eq:pS}
\end{align}
with the last line following from a straightforward calculation using Eq.~\eqref{eq:rhoE} and $\mathcal{U}=U^{\otimes 4}$. 

Experimentally, this scenario can be realized, for example, via the scattering of four photons in a four-mode beam splitter, which implements the unitary~\eqref{eq:U} (e.g. as in \cite{Ehrhardt-EC-2020}), followed by a measurement of the number of photons in the output ports using a photon-number-resolving detection. In this setting, the optical modes of the beam splitter correspond to the photons' external (i.e. dynamical) degrees of freedom, while all other degrees of freedom, such as the photons' polarizations or frequency spectra, are lumped together as their internal degrees of freedom.

\subsection{Classifying the output events}
In view of the tensor power in Eq.~\eqref{eq:U}, the hypercube unitary $U$ can be seen as the two-dimensional generalization of a balanced two-mode coupler (or beam splitter). As illustrated in Fig.~\ref{fig:HC}(b), the Hamiltonian's coupling structure between the modes gives rise to a square-shaped graph \cite{Dittel-MB-2017} whose symmetries are inscribed into $U$. In particular, $U$ exhibits three symmetries $\mathpzc{S} \in\{1,2,3\}$. In the graph structure in Fig.~\ref{fig:HC}(b), these symmetries are associated with the mode permutations \footnote{Note that in~\eqref{eq:sigma} we name the permutations $\sigma$ since (different to $\kappa$ and $\pi$) they act on modes rather than particles.}
\begin{align}\label{eq:sigma}
\begin{split}
\sigma^{(1)}&=(1~3)(2~4),\\
\sigma^{(2)}&=(1~2)(3~4),\\
\sigma^{(3)}&=(1~4)(2~3),
\end{split}
\end{align}
which, together with the identity $\epsilon$, form the normal Klein four-subgroup $\mathrm{K}_4=\{\epsilon, \sigma^{(1)}, \sigma^{(2)}, \sigma^{(3)}\}$ of $\mathrm{S}_4$ \cite{Baumslag-SO-1968}. Note that the symmetries $\mathpzc{S}$ associated with permutations $\sigma^{(\mathpzc{S})}$ are noted as superscripts, since, as defined at the beginning of Sec.~\ref{sec:quant}, we use subscripts to label the cycles of permutations, e.g. $\sigma^{(1)}_2=(2~4)$. 

By Eq.~\eqref{eq:pS}, the hypercube's symmetries also manifest in the output statistics for partially distinguishable particles. In particular, as proven in Appendix~\ref{app:pSsigma-pS}, we find that all output events $\vec{S}$ differing by mode permutations $\sigma^{(\mathpzc{S})}$ of the hypercube symmetries from Eq.~\eqref{eq:sigma} appear with equal transition probability~\eqref{eq:pS}, i.e.
\begin{align}\label{eq:pSsigma-pS}
p_{\vec{S}_{\sigma} }= p_{\vec{S}},
\end{align}
with $\vec{S}_{\sigma}=(S_{\sigma(1)}, S_{\sigma(2)} , S_{\sigma(3)}, S_{\sigma(4)})$, and $\sigma \in \mathrm{K}_4$. This multiplicity suggests to collect all output events $\vec{S}$ which appear with equal transition probabilities. Therefore, let us define the following \emph{classes} of output events:
\begin{align}\label{eq:FclassI}
\begin{split}
\mathcal{F}_1^\mathrm{I}=\{(3,1,0,0),(1,3,0,0),(0,0,3,1),(0,0,1,3)\}, \\
\mathcal{F}_2^\mathrm{I}=\{(3,0,1,0),(0,3,0,1),(1,0,3,0),(0,1,0,3)\}, \\
\mathcal{F}_3^\mathrm{I}=\{(3,0,0,1),(0,3,1,0),(0,1,3,0),(1,0,0,3)\},
\end{split}
\end{align}
and
\begin{align}\label{eq:FclassII}
\begin{split}
\mathcal{F}_1^\mathrm{II}=\{(2,0,1,1),(0,2,1,1),(1,1,2,0),(1,1,0,2)\}, \\
\mathcal{F}_2^\mathrm{II}=\{(2,1,0,1),(1,2,1,0),(0,1,2,1),(1,0,1,2)\}, \\
\mathcal{F}_3^\mathrm{II}=\{(2,1,1,0),(1,2,0,1),(1,0,2,1),(0,1,1,2)\}.
\end{split}
\end{align}
The classes $\mathcal{F}_1^\mathrm{I(II)}$, $\mathcal{F}_2^\mathrm{I(II)}$, and $\mathcal{F}_3^\mathrm{I(II)}$ contain those output events $\vec{S}$ with an even number of particles in the subset of modes $\{1,2\}$, $\{1,3\}$, and $\{1,4\}$, respectively, with events in $\mathcal{F}_\mathpzc{S}^\mathrm{I}$ ($\mathcal{F}_\mathpzc{S}^\mathrm{II}$) corresponding to permutations of $\vec{S}=(3,1,0,0)$ ($\vec{S}=(2,1,1,0)$). In the following these events are called events of \emph{type} $\mathrm{I}$ ($\mathrm{II}$). Considering the permutations $\sigma^{(\mathpzc{S})}$ from Eq.~\eqref{eq:sigma} and the symmetry~\eqref{eq:pSsigma-pS} of the output statistics, one can easily see that all output events in the same class must appear with equal transition probability. On the basis of Eq.~\eqref{eq:pSsigma-pS}, we also classify the remaining output events,
\begin{align}
\mathcal{A}_\mathrm{A}&=\{(1,1,1,1)\}, \nonumber \\
\mathcal{A}_\mathrm{B}&=\{(4,0,0,0),(0,4,0,0),(0,0,4,0),(0,0,0,4)\}, \nonumber \\
\mathcal{A}_1&=\{(2,2,0,0),(0,0,2,2)\}, \label{eq:A} \\
\mathcal{A}_2&=\{(2,0,2,0),(0,2,0,2)\}, \nonumber \\
\mathcal{A}_3&=\{(2,0,0,2),(0,2,2,0)\} .\nonumber
\end{align}
with $\mathcal{A}_\mathrm{A}$ containing the anti-bunched ($\mathrm{A}$) event, $\mathcal{A}_\mathrm{B}$ all bunched ($\mathrm{B}$) events, and $\mathcal{A}_1$, $\mathcal{A}_2$, and $\mathcal{A}_3$ those events with either two or no particles in each mode of the subset $\{1,2\}$, $\{1,3\}$, and $\{1,4\}$, respectively. In total, we see that the output statistics contain at most eleven distinct transition probabilities which are specified hereafter. Note that in the case of strictly indistinguishable bosons, all output events listed in~\eqref{eq:FclassI} and~\eqref{eq:FclassII} are forbidden ($\mathcal{F}$) due to totally destructive many-particle interference \cite{Dittel-MB-2017,Viggianiello-EG-2018,Dittel-TD1-2018,Dittel-TD2-2018,Dittel-AI-2019}, while those events listed in~\eqref{eq:A} are allowed ($\mathcal{A}$), i.e. appear with a finite transition probability.

\subsection{Output statistics}
Let us now determine the transition probabilities~\eqref{eq:pS} for the output events from all eleven classes~\eqref{eq:FclassI}-\eqref{eq:A}, and express them in terms of the traces~\eqref{eq:CollPhase}, which, by Eq.~\eqref{eq:rhoElementTr}, enter~\eqref{eq:pS} through the off-diagonal elements of the reduced external state $\rho_\mathrm{E}$. First  consider the classes $\mathcal{F}_\mathpzc{S}^{\mathrm{I(II)}}$: As we show in Appendix~\ref{app:PF}, the transition probabilities of output events $\vec{S}$ in these classes read [recall that the upper (lower) sign corresponds to the case of bosons (fermions)]
\begin{widetext}
\begin{align}\label{eq:pF}
\begin{split}
p_{\mathcal{F}_1^\mathrm{I}}&=\frac{1}{64} \left[1\pm T_{(13)} \pm T_{(24)} + T_{(13)}T_{(24)}- T_{(12)}T_{(34)} -T_{(14)}T_{(23)}\mp 2 T_{(1234)} \cos\varphi_{(1234)} \right],\\
p_{\mathcal{F}_2^\mathrm{I}}&=\frac{1}{64} \left[1\pm T_{(12)} \pm T_{(34)} + T_{(12)}T_{(34)}- T_{(13)}T_{(24)} -T_{(14)}T_{(23)}\mp2 T_{(1324)} \cos\varphi_{(1324)}\right],\\
p_{\mathcal{F}_3^\mathrm{I}}&=\frac{1}{64} \left[1\pm T_{(14)} \pm T_{(23)} + T_{(14)}T_{(23)}- T_{(12)}T_{(34)} -T_{(13)}T_{(24)}\mp 2 T_{(1243)} \cos\varphi_{(1243)}\right],\\
p_{\mathcal{F}_1^\mathrm{II}}&=\frac{1}{64} \left[3\mp T_{(13)} \mp T_{(24)} +3 T_{(13)}T_{(24)}-3 T_{(12)}T_{(34)} -3T_{(14)}T_{(23)}\pm 2T_{(1234)} \cos\varphi_{(1234)} \right],\\
p_{\mathcal{F}_2^\mathrm{II}}&=\frac{1}{64} \left[3\mp T_{(12)} \mp T_{(34)} +3 T_{(12)}T_{(34)}-3 T_{(13)}T_{(24)} -3T_{(14)}T_{(23)}\pm2T_{(1324)} \cos\varphi_{(1324)}\right],\\
p_{\mathcal{F}_3^\mathrm{II}}&=\frac{1}{64} \left[3\mp T_{(14)} \mp T_{(23)} +3 T_{(14)}T_{(23)}-3 T_{(12)}T_{(34)} -3T_{(13)}T_{(24)}\pm2T_{(1243)} \cos\varphi_{(1243)}\right].
\end{split}
\end{align}
\end{widetext}
On the other hand, from similar considerations as in Appendix~\ref{app:PF}, one can show that the transition probabilities of output events $\vec{S}$ in the classes~\eqref{eq:A} can be written as
\begin{widetext}
\begin{align}\label{eq:pA}
\begin{split}
p_{\mathcal{A}_\mathrm{A}}&= \frac{1}{32}\left[ 3\mp \mathcal{I}_{(1,1,2)} + \mathcal{I}_{(1,3)}+3\mathcal{I}_{(2,2)} \mp \mathcal{I}_{(4)} \right],\\
p_{\mathcal{A}_\mathrm{B}}&= \frac{1}{256}\left[ 1 \pm\mathcal{I}_{(1,1,2)} + \mathcal{I}_{(1,3)}+\mathcal{I}_{(2,2)} \pm\mathcal{I}_{(4)} \right],\\
p_{\mathcal{A}_1}&= \frac{1}{128}\left[ 3\mp \mathcal{I}_{(1,1,2)} - \mathcal{I}_{(1,3)}+3\mathcal{I}_{(2,2)} \mp\mathcal{I}_{(4)} \pm 4 T_{(13)} \pm 4 T_{(24)}\pm 8T_{(1234)} \cos\varphi_{(1234)} \right],\\
p_{\mathcal{A}_2}&= \frac{1}{128}\left[ 3\mp\mathcal{I}_{(1,1,2)} - \mathcal{I}_{(1,3)}+3\mathcal{I}_{(2,2)} \mp\mathcal{I}_{(4)} \pm 4 T_{(12)} \pm 4T_{(34)}\pm 8T_{(1324)} \cos\varphi_{(1324)} \right],\\
p_{\mathcal{A}_3}&= \frac{1}{128}\left[ 3\mp\mathcal{I}_{(1,1,2)} - \mathcal{I}_{(1,3)}+3\mathcal{I}_{(2,2)} \mp\mathcal{I}_{(4)} \pm 4 T_{(14)} \pm 4T_{(23)}\pm 8T_{(1243)} \cos\varphi_{(1243)} \right],
\end{split}
\end{align}
\end{widetext}
with the indistinguishability quantifiers $\mathcal{I}_L$ from Eq.~\eqref{eq:Dall}. Equations~\eqref{eq:pF} and~\eqref{eq:pA} clearly demonstrate that, due to the linearity of quantum mechanics, the off-diagonal elements~\eqref{eq:rhoElementTr} associated with the indistinguishability of different particle labellings [see Sec.~\ref{sec:quant}] enter linearly in the output statistics. In the next section we proceed with analyzing these transition probabilities.

\section{Characterization of the particles' indistinguishability}\label{sec:Chraracterizing}
Let us recall that neither the unitary transformation imparted upon the particles, nor the subsequent detection, considered in the previous section, act on the particles' internal degrees of freedom. However, as a direct consequence of many-particle interference, the resulting output statistics [see Eqs.~\eqref{eq:pF} and~\eqref{eq:pA}] strongly depends on the particles' indistinguishability. Moreover, the symmetry inherent in the hypercube unitary~\eqref{eq:U} also led to rather simple and well-structured dependencies of the transition probabilities on the particles' indistinguishability. As we show in the following, this, in turn, allows us to extract information about the particles' indistinguishability via simple, mostly linear combinations of the transition probabilities.

\subsection{Direct analysis}
With the output statistics at hand, we now define combinations of the transition probabilities from Eqs.~\eqref{eq:pF} and~\eqref{eq:pA}, which allow us to extract the indistinguishability quantifier discussed in Sec.~\ref{sec:quant}. For reasons of compactness we choose $\mathpzc{S} \neq \mathpzc{S}' \neq \mathpzc{S}'' \neq \mathpzc{S}$, and use the cycle representation [see Sec.~\ref{sec:quant}] of permutations $\sigma^{(\mathpzc{S})}$ from Eq.~\eqref{eq:sigma}, e.g. $\sigma^{(3)}=(1~4)(2~3)$ with cycles $\sigma^{(3)}_1=(1~4)$ and $\sigma^{(3)}_2=(2~3)$, such that $T_{(14)}\equiv T_{\sigma^{(3)}_1}$ and $T_{(23)}\equiv T_{\sigma^{(3)}_2}$.

Let us first define combinations of transition probabilities which are useful for our analysis of the particles' indistinguishability. To this end, we add the probabilities for output events in classes $\mathcal{F}_\mathpzc{S}^\mathrm{I}$ and $\mathcal{F}_\mathpzc{S}^\mathrm{II}$, with equal symmetry $\mathpzc{S}\in\{1,2,3\}$, 
\begin{align}\label{eq:PFS}
P_{\mathcal{F}_\mathpzc{S}}:=p_{\mathcal{F}_\mathpzc{S}^\mathrm{I}}+p_{\mathcal{F}_\mathpzc{S}^\mathrm{II}}.
\end{align}
Combinations thereof lead to products of two-particle indistinguishabilities~\eqref{eq:W2p} via
\begin{align}\label{eq:TT}
\mathcal{P}_\mathpzc{S}&:=1-8\left[P_{\mathcal{F}_{\mathpzc{S}'}} +P_{\mathcal{F}_{\mathpzc{S}''}}\right]=T_{\sigma^{(\mathpzc{S})}_1} T_{\sigma^{(\mathpzc{S})}_2},
\end{align}
and the sum of $T_{\sigma^{(\mathpzc{S})}_1}$ and $T_{\sigma^{(\mathpzc{S})}_2}$ can be constructed as
\begin{align}
\mathcal{Q}_\mathpzc{S}&:=16 p_{\mathcal{A}_\mathrm{B}} + 8 p_{\mathcal{A}_\mathpzc{S}} + 16 p_{\mathcal{F}_\mathpzc{S}^\mathrm{I}} + 4\left[ P_{\mathcal{F}_{\mathpzc{S}'}} +P_{\mathcal{F}_{\mathpzc{S}''}} \right]-1\nonumber\\
&=\pm \frac{1}{2}\left[ T_{\sigma^{(\mathpzc{S})}_1} + T_{\sigma^{(\mathpzc{S})}_2}\right].\label{eq:Q1}
\end{align}
Note that both $\mathcal{P}_\mathpzc{S}$ and $\mathcal{Q}_\mathpzc{S}$ are linear in the transition probabilities from Eqs.~\eqref{eq:pF} and~\eqref{eq:pA}.

We now follow our discussion in Sec.~\ref{sec:quant}, and show how to extract all indistinguishability measures defined there. First let us consider the indistinguishability quantifier $\mathcal{I}_L$ for particle labellings with cycle structure $L$. Equation~\eqref{eq:Dall} together with~\eqref{eq:TT} and~\eqref{eq:Q1} allows the immediate inference of $\mathcal{I}_{(1,1,2)}$ and $\mathcal{I}_{(2,2)}$ from the $\mathcal{P}_\mathpzc{S}$ and $\mathcal{Q}_\mathpzc{S}$, through
\begin{align}\label{eq:I112-p}
\mathcal{I}_{(1,1,2)}=\pm 2\left[ \mathcal{Q}_1+\mathcal{Q}_2+\mathcal{Q}_3  \right],
\end{align}
and
\begin{align}\label{eq:I22-p}
\mathcal{I}_{(2,2)}=\mathcal{P}_1+\mathcal{P}_2+\mathcal{P}_3.
\end{align}
Furthermore, Eqs.~\eqref{eq:pF} and~\eqref{eq:pA}, together with~\eqref{eq:Dall} and~\eqref{eq:PFS}, lead to the following expressions for $\mathcal{I}_{(1,3)}$ and $\mathcal{I}_{(4)}$,
\begin{align}\label{eq:I13-p}
\mathcal{I}_{(1,3)}=128 p_{\mathcal{A}_\mathrm{B}} + 16 p_{\mathcal{A}_\mathrm{A}} +32\left[P_{\mathcal{F}_1} +P_{\mathcal{F}_2}+P_{\mathcal{F}_3}\right]-8,
\end{align}
and
\begin{align}\label{eq:I4-p}
\begin{split}
\mathcal{I}_{(4)}&=\pm \Big(96 p_{\mathcal{A}_\mathrm{B}} +16\left[ p_{\mathcal{A}_1} +p_{\mathcal{A}_2} +p_{\mathcal{A}_3} \right]  \\
 &+ 32 \left[ p_{\mathcal{F}_1^\mathrm{II}} +p_{\mathcal{F}_2^\mathrm{II}}+p_{\mathcal{F}_3^\mathrm{II}} \right] -6\Big),
 \end{split}
\end{align}
such that \emph{all} $\mathcal{I}_L$ can be directly distilled from the output statistics. Once all $\mathcal{I}_L$ are determined, also the $N$-particle indistinguishability quantifiers $\braket{\Pi_\mathrm{B(F)}}_{N\mathrm{p}}$ are, by Eqs.~\eqref{eq:W4p}-\eqref{eq:Wa2p}. Indeed, for bosons, $\braket{\Pi_\mathrm{B(F)}}_{4\mathrm{p}}$ is even directly read off the expression for $p_{\mathcal{A}_\mathrm{B}}$ in~\eqref{eq:pA},
\begin{align}\label{eq:W4pch}
\braket{\Pi_\mathrm{B(F)}}_\mathrm{4p}=\frac{32}{3} \ p_{\mathcal{A}_\mathrm{B}}.
\end{align}

Let us note that we can even resolve all individual terms of the four-particle contributions in $\mathcal{I}_{(4)}$ [see Eq.~\eqref{eq:Dall}]. In particular, we find
\begin{align}\label{eq:T1234Statistics}
\begin{split}
\pm T_{(1234)} \cos \varphi_{(1234)} &= 16p_{\mathcal{A}_\mathrm{B}} +8p_{\mathcal{A}_1} -8\left[p_{\mathcal{F}_1^\mathrm{I}} -p_{\mathcal{F}_1^\mathrm{II}}\right] \\
&+4\left[ P_{\mathcal{F}_2} +P_{\mathcal{F}_3} \right]-1,\\
\pm T_{(1324)} \cos \varphi_{(1324)} &= 16p_{\mathcal{A}_\mathrm{B}} +8p_{\mathcal{A}_2} -8\left[p_{\mathcal{F}_2^\mathrm{I}} -p_{\mathcal{F}_2^\mathrm{II}}\right] \\
&+4\left[ P_{\mathcal{F}_1} +P_{\mathcal{F}_3} \right]-1,\\
\pm T_{(1243)} \cos \varphi_{(1243)} &= 16p_{\mathcal{A}_\mathrm{B}} +8p_{\mathcal{A}_3} -8\left[p_{\mathcal{F}_3^\mathrm{I}} -p_{\mathcal{F}_3^\mathrm{II}}\right] \\
&+4\left[ P_{\mathcal{F}_1} +P_{\mathcal{F}_2} \right]-1.
\end{split}
\end{align}

In the case of three specific particles $\{\alpha,\beta,\gamma\}$, let us stress that the quantifier $\braket{\Pi_\mathrm{B(F)}}_{(\alpha,\beta,\gamma)}$ from Eq.~\eqref{eq:W3p} contains only a single term of the form $T_{(\alpha\beta\gamma)} \cos \varphi_{(\alpha\beta\gamma)}$. In the output statistics given in~\eqref{eq:pA}, these terms always occur together with all other three-particle terms of the same form, through $\mathcal{I}_{(1,3)}$ [see Eq.~\eqref{eq:Dall}]. Consequently, the three-particle indistinguishability $\braket{\Pi_\mathrm{B(F)}}_{(\alpha,\beta,\gamma)}$ cannot be extracted directly from the output statistics. However, we will see in Sec.~\ref{sec:symbreak} below that this can be achieved with the help of an additional, reasonable assumption.

The two-particle indistinguishabilities~\eqref{eq:W2p} are all of the form $[1+ T_{\sigma^{(\mathpzc{S})}_j}]/2$, with $j\in\{1,2\}$, by virtue of~\eqref{eq:sigma}. Equations~\eqref{eq:TT} and~\eqref{eq:Q1} show that these quantities are given by combinations of $\mathcal{P}_\mathpzc{S}$ and $\mathcal{Q}_\mathpzc{S}$. In particular, for $T_{\sigma^{(\mathpzc{S})}_1}$ and $T_{\sigma^{(\mathpzc{S})}_2}$ we obtain two solutions which apply depending on their relative magnitude:
\begin{align}\label{eq:Tsqrt1}
\pm T_{\sigma^{(\mathpzc{S})}_1}&=\begin{cases}   \mathcal{Q}_\mathpzc{S}\pm \sqrt{ \mathcal{Q}_\mathpzc{S}^2 -\mathcal{P}_\mathpzc{S}} \quad \text{if}\quad T_{\sigma^{(\mathpzc{S})}_1}\geq T_{\sigma^{(\mathpzc{S})}_2}\\ 
\mathcal{Q}_\mathpzc{S}\mp\sqrt{ \mathcal{Q}_\mathpzc{S}^2 -\mathcal{P}_\mathpzc{S}} \quad \text{if}\quad T_{\sigma^{(\mathpzc{S})}_1}< T_{\sigma^{(\mathpzc{S})}_2}, \end{cases}
\end{align}
and
\begin{align}\label{eq:Tsqrt2}
\pm T_{\sigma^{(\mathpzc{S})}_2}&=\begin{cases}   \mathcal{Q}_\mathpzc{S}\mp\sqrt{ \mathcal{Q}_\mathpzc{S}^2 -\mathcal{P}_\mathpzc{S}} \quad \text{if}\quad T_{\sigma^{(\mathpzc{S})}_1}\geq T_{\sigma^{(\mathpzc{S})}_2}\\ 
\mathcal{Q}_\mathpzc{S}\pm\sqrt{ \mathcal{Q}_\mathpzc{S}^2 -\mathcal{P}_\mathpzc{S}} \quad \text{if}\quad T_{\sigma^{(\mathpzc{S})}_1}< T_{\sigma^{(\mathpzc{S})}_2}. \end{cases}
\end{align}
That is to say, we can extract \emph{all} two-particle indistinguishabilities $\braket{\Pi_\mathrm{B(F)}}_{(\alpha,\beta)}$ from Eq.~\eqref{eq:W2p}, except for a discrimination between $T_{\sigma^{(\mathpzc{S})}_1}$ and $T_{\sigma^{(\mathpzc{S})}_2}$, for all $\mathpzc{S}\in\{1,2,3\}$. This remaining ambiguity is a natural consequence of the transition probabilities' symmetry~\eqref{eq:pSsigma-pS}, and manifests in the fact that these are invariant under exchange of $T_{\sigma^{(\mathpzc{S})}_1}$ and $T_{\sigma^{(\mathpzc{S})}_2}$ [see Eqs.~\eqref{eq:pF} and~\eqref{eq:pA}]. Hence, there is no way to distinguish the two cases in~\eqref{eq:Tsqrt1} and~\eqref{eq:Tsqrt2} from the output statistics. However, we will see in Sec.~\ref{sec:symbreak} below that such distinction becomes possible by the same argument which will also allow to extract the three-particle indistinguishability quantifiers $\braket{\Pi_\mathrm{B(F)}}_{(\alpha,\beta,\gamma)}$.

We thus have shown that all contributions~\eqref{eq:trcos} to the indistinguishability quantifiers discussed in Sec.~\ref{sec:quant} follow directly from the counting statistics on output -- \emph{except} for three-particle contributions of the form $T_{(\alpha\beta\gamma)} \cos \varphi_{(\alpha\beta\gamma)}$, and the distinction of the two-particle contributions $T_{\sigma^{(\mathpzc{S})}_1}$ and $T_{\sigma^{(\mathpzc{S})}_2}$. These remaining issues will be dealt with now.

\subsection{Characterizing the exceptions}\label{sec:symbreak}
Let us assume that each of the four particles can be made distinguishable from the remaining three. Since the hypercube unitary~\eqref{eq:U} describes a non-interacting evolution, this is equivalent to the ability of dropping this particle (since it will not contribute to the particles' mutual interference), and to combine the counting statistics of the remaining particles with that of the single dropped particle.

Now, let us take the two-particle indistinguishabilities $T_{(\alpha \beta)}=T_{\sigma^{(\mathpzc{S})}_1}$ and $T_{(\gamma \delta)}=T_{\sigma^{(\mathpzc{S})}_2}$ from Eqs.~\eqref{eq:Tsqrt1} and~\eqref{eq:Tsqrt2}, and make particle $\delta$ fully distinguishable from all other particles. Thus, we have $T_{\sigma^{(\mathpzc{S})}_2}=0$, while $T_{\sigma^{(\mathpzc{S})}_1}$ remains unaffected and can be extracted from Eq.~\eqref{eq:Q1}:
\begin{align} \label{eq:Tabch}
T_{\sigma^{(\mathpzc{S})}_1}=\pm 2 \mathcal{Q}_\mathpzc{S}.
\end{align}
This allows us to distinguish the cases in Eqs.~\eqref{eq:Tsqrt1} and~\eqref{eq:Tsqrt2}. Moreover, in this case, the four-particle indistinguishability $\braket{\Pi_\mathrm{B(F)}}_\mathrm{4p}$ from Eq.~\eqref{eq:W4p} reduces to the three-particle indistinguishability $\braket{\Pi_\mathrm{B(F)}}_{(\alpha,\beta,\gamma)}$ from Eq.~\eqref{eq:W3p} up to a constant factor, which, therefore, by Eqs.~\eqref{eq:W4p}, \eqref{eq:W3p}, and~\eqref{eq:pA}, is given as
\begin{align}\label{eq:W3pch}
\braket{\Pi_\mathrm{B(F)}}_{(\alpha,\beta,\gamma)}= \frac{128}{3}\ p_{\mathcal{A}_\mathrm{B}} 
\end{align}
for bosons, and for fermions we find
\begin{align}\label{eq:W3pchF}
\braket{\Pi_\mathrm{B(F)}}_{(\alpha,\beta,\gamma)}= \frac{1}{3} \left(16 p_{\mathcal{A}_\mathrm{A}} -1 \right) .
\end{align}
In summary, by making one particle fully distinguishable from all others, it does not contribute to the particles' mutual interference, and, hence, the output statistics change. Given this option we can extract \emph{all} individual contributions~\eqref{eq:trcos} to the indistinguishability quantifiers of all subsets of particles.

\section{State reconstruction for pure internal states}\label{sec:Pure-internal-states}
While in the previous section we showed how to extract all $N$-particle contributions~\eqref{eq:trcos} to the indistinguishability quantifiers defined in Sec.~\ref{sec:quant}, so far the exact values of the collective phases $\varphi_{(\alpha \beta \cdots \gamma)}$ in Eq.~\eqref{eq:CollPhase} remain unspecified. To the best of our knowledge, for mixed internal states~\eqref{eq:rho} the characterization of these phases goes beyond the capabilities of the here presented scheme. However, for the case of pure internal states~\eqref{eq:rho} there are explicit interrelations between different collective phases \cite{Shchesnovich-CP-2018} defined in Eq.~\eqref{eq:CollPhase}, which allow their characterization with the help of the output statistics~\eqref{eq:pF} and~\eqref{eq:pA}, and, ultimately, a reconstruction of the reduced external many-body density operator $\rho_\mathrm{E}$ from Eq.~\eqref{eq:rhoE}. To elaborate on this, we discuss the relations between the collective phases in the following, and continue in Sec.~\ref{sec:pureRecon} with their extraction from the counting statistics and the reconstruction of $\rho_\mathrm{E}$.

\subsection{Collective phases for pure internal states}\label{sec:PureColPhase}
Let us consider pure internal states, with the internal density operator of the $\alpha$th particle reading $\rho_\alpha=\ketbra{\phi_\alpha}{\phi_\alpha}$. In this case Eq.~\eqref{eq:CollPhase} becomes
\begin{align}\label{eq:TrabyPure}
\begin{split}
\tr{\rho_\alpha \rho_\beta \cdots \rho_\gamma} &=\bracket{\phi_\gamma}{\phi_\alpha}\bracket{\phi_\alpha}{\phi_\beta} \bra{\phi_\beta}\cdots \ket{\phi_\gamma},\\
&= \abs{\bracket{\phi_\gamma}{\phi_\alpha}\bracket{\phi_\alpha}{\phi_\beta} \bra{\phi_\beta}\cdots \ket{\phi_\gamma}} e^{\mathrm{i} \varphi_{(\alpha \beta \cdots \gamma)}}.
\end{split}
\end{align}
Utilizing
\begin{align}\label{eq:VarphiAKB}
\bracket{\phi_\alpha}{\phi_\beta}=\abs{\bracket{\phi_\alpha}{\phi_\beta}} e^{\mathrm{i}\vartheta_{\alpha,\beta}}, 
\end{align}
with $\vartheta_{\alpha,\beta}=-\vartheta_{\beta,\alpha}$, the collective phases [see Eq.~\eqref{eq:CollPhase}] become \cite{Shchesnovich-CP-2018} 
\begin{align}\label{eq:CollPhasePure}
\begin{split}
\varphi_{(\alpha\beta)}&=\vartheta_{\alpha,\beta}+\vartheta_{\beta,\alpha}=0\\
\varphi_{(\alpha\beta\gamma)}&=\vartheta_{\alpha,\beta}+\vartheta_{\beta,\gamma}+\vartheta_{\gamma,\alpha},\\
\varphi_{(\alpha\beta\gamma\delta)}&=\vartheta_{\alpha,\beta}+\vartheta_{\beta,\gamma}+ \vartheta_{\gamma,\delta}+\vartheta_{\delta,\alpha}.
\end{split}
\end{align}
Note the difference between the collective phases $\varphi_{(\alpha\beta\cdots\gamma)}$ from Eq.~\eqref{eq:CollPhase} and $\vartheta_{\alpha,\beta}$ from Eq.~\eqref{eq:VarphiAKB}, in particular note that $\varphi_{(\alpha\beta)}=0$ for all $\alpha,\beta\in\{1,2,3,4\}$ while $\vartheta_{\alpha,\beta}$ is not necessarily zero.

The expressions in~\eqref{eq:CollPhasePure} govern the interrelation between different collective phases for pure internal states. However, by Eq.~\eqref{eq:TrabyPure}, vanishing overlaps $\bracket{\phi_\alpha}{\phi_\beta}$ of the particles' internal states can lead to the absence of certain collective phases. We therefore need to discuss the relation between non-vanishing collective phases and the number of orthogonal internal states. The different cases that arise are illustrated by graphs in Fig.~\ref{fig:COLP}, where vertices and edges correspond to particles and their overlaps, respectively. Accordingly, by Eqs.~\eqref{eq:CollPhase} and~\eqref{eq:TrabyPure}, each possible closed loop in the graph representation gives rise to two collective phases corresponding to inverse cycles $\kappa_j$ and $\kappa_j^{-1}$ \cite{Shchesnovich-CP-2018}, e.g. in Fig.~\ref{fig:COLP}(e) the clockwise loop is associated with $\varphi_{(\beta\gamma\delta)}=\mathrm{arg}(\bracket{\phi_\beta}{\phi_\gamma}\bracket{\phi_\gamma}{\phi_\delta}\bracket{\phi_\delta}{\phi_\beta})$, and the anti-clockwise loop with $\varphi_{(\beta\delta\gamma)}$.

We start with non-orthogonal internal states as represented by a fully connected graph shown in Fig.~\ref{fig:COLP}(a). In this case, collective phases between all possible subsets of particles arise. In \cite{Shchesnovich-CP-2018} it was shown that all these phases can be obtained from linear combinations of only three independent triad phases (i.e. collective phases involving three particles). This can be seen as follows: first recall from our derivation of~\eqref{eq:trcos} that for inverse cycles we have $\tr{\rho_\alpha \rho_\beta \cdots \rho_\gamma  }=\tr{\rho_\gamma \cdots \rho_\beta \rho_\alpha  }^*$, such that
\begin{align}\label{eq:varphi-}
\varphi_{(\alpha \beta \cdots \gamma)}=-\varphi_{(\gamma \cdots \beta\alpha )}.
\end{align}
Together with~\eqref{eq:CollPhasePure}, a short calculation then reveals
\begin{align}\label{eq:varphi3}
\varphi_{(123)}+\varphi_{(134)} = \varphi_{(124)}+\varphi_{(234)},
\end{align}
and
\begin{align}\label{eq:varphi4}
\begin{split}
\varphi_{(1234)}&=\varphi_{(123)}+\varphi_{(134)} ,\\
\varphi_{(1423)}&=\varphi_{(123)}+\varphi_{(142)} ,\\
\varphi_{(1342)}&=\varphi_{(134)}+\varphi_{(142)}.
\end{split}
\end{align}
Note that the decomposition~\eqref{eq:varphi4} can be illustrated via the graph representation in Fig.~\ref{fig:COLP}: Each closed four-particle loop with collective phase $\varphi_{(\beta\gamma\delta\alpha)}$ [e.g. see the loop $\beta \rightarrow \gamma \rightarrow \delta \rightarrow \alpha \rightarrow \beta$ in Fig.~\ref{fig:COLP}(b)] can be built from two closed three-particle loops (e.g. $\beta \rightarrow \gamma \rightarrow \delta \rightarrow \beta$ and $\beta \rightarrow \delta \rightarrow \alpha \rightarrow \beta$ with collective phases $\varphi_{(\beta\gamma\delta)}$ and $\varphi_{(\beta\delta\alpha)}$, respectively) whose overlapping edges (e.g. between particle $\beta$ and $\delta$) cancel due to inverse loop directions. Altogether, given three independent triad phases, we see that, with the help of Eqs.~\eqref{eq:varphi-}-\eqref{eq:varphi4}, we obtain all of the -- in total -- eight and six collective phases involving three and four particles, corresponding to all cycles of length three and four, respectively. 

\begin{figure}[t]
\centering
\includegraphics[width=\linewidth]{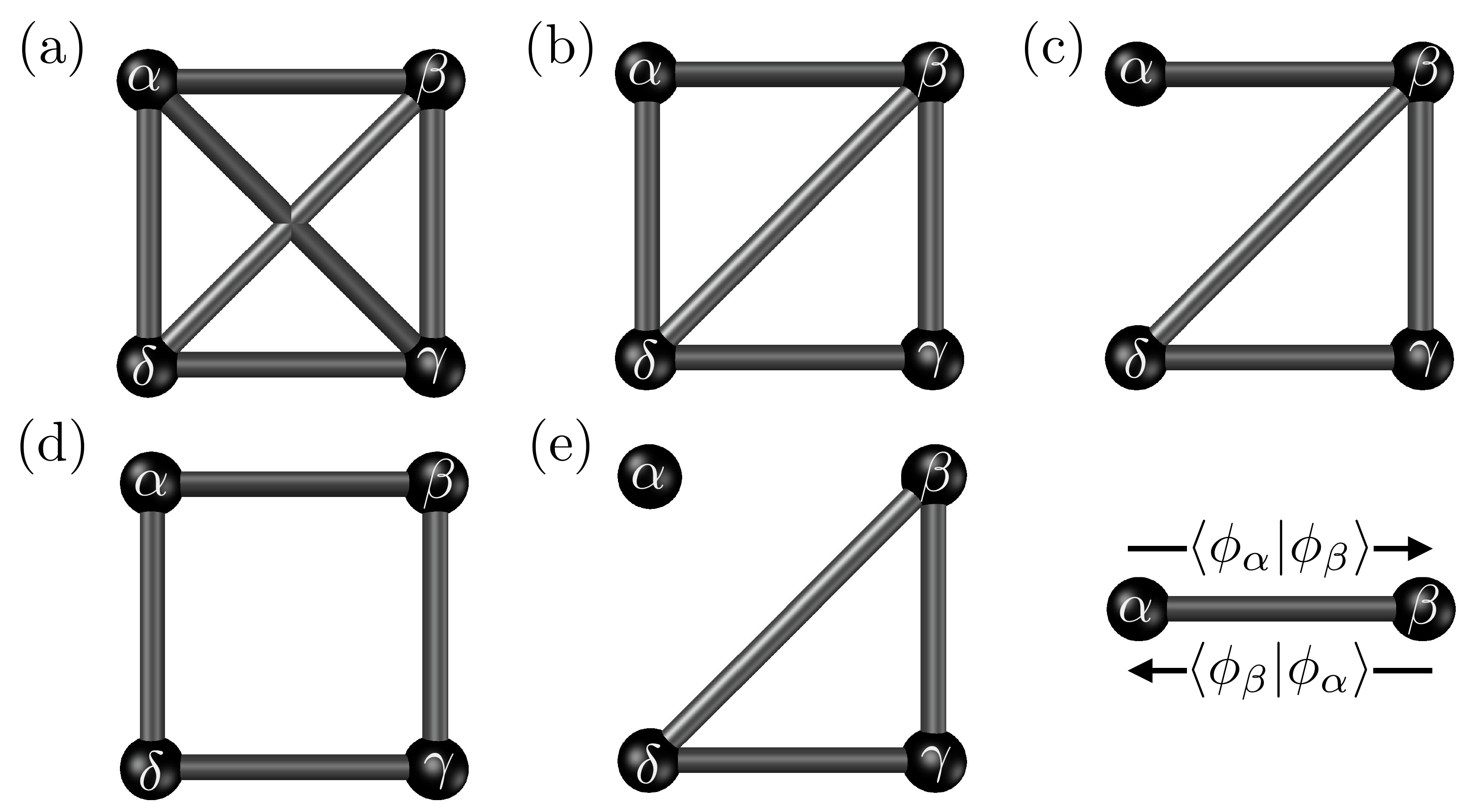}
\caption{Graph representations of the particles' overlaps. Particles $\alpha$, $\beta$, $\gamma$, and $\delta$ are illustrated by the graphs' vertices, and their non-vanishing overlaps by edges. As illustrated in the bottom right graph, when moving from vertex $\alpha$ to $\beta$ ($\beta$ to $\alpha$) the edge corresponds to $\langle \phi_\alpha|\phi_\beta\rangle$ ($\langle \phi_\beta|\phi_\alpha\rangle$). All sets of vertices connected by closed loops give rise to collective phases, which appear for (a) non-orthogonal internal states, (b) $\langle \phi_\alpha|\phi_\gamma\rangle=0$, (c) $\langle \phi_\alpha|\phi_\gamma\rangle=\langle \phi_\alpha|\phi_\delta\rangle=0$, (d) $\langle \phi_\alpha|\phi_\gamma\rangle=\langle \phi_\beta|\phi_\delta\rangle=0$, and (e) $\langle \phi_\alpha|\phi_\gamma\rangle=\langle \phi_\alpha|\phi_\delta\rangle=\langle \phi_\alpha|\phi_\beta\rangle=0$. }
\label{fig:COLP}
\end{figure}

Next let us suppose that particles $\alpha$ and $\gamma$ are in orthogonal internal states, i.e., $\bracket{\phi_\alpha}{\phi_\gamma}=0$. Accordingly, in the corresponding graph in Fig.~\ref{fig:COLP}(b) there is a missing edge between the vertices associated with these particles. In this case, we are left with four triad phases, $\varphi_{(\beta \delta \alpha)}$, $\varphi_{(\beta \gamma \delta)}$, and the phases with inverse cycles $\varphi_{(\beta \alpha \delta)}$, and $\varphi_{(\beta \delta \gamma)}$. Moreover, there are two collective phases involving four particles, $\varphi_{(\beta \gamma \delta \alpha)}$ and the phase corresponding to the inverse cycle, $\varphi_{(\beta \alpha \delta \gamma)}$. All other collective phases are absent. Again, using Eqs.~\eqref{eq:CollPhasePure} and~\eqref{eq:varphi-}, a short calculation reveals
\begin{align}\label{eq:ColPhasMisEdge}
\varphi_{(\beta \gamma \delta \alpha)}=\varphi_{(\beta \gamma \delta)}+\varphi_{(\beta \delta \alpha)}.
\end{align}
Thus, similar to Eq.~\eqref{eq:varphi4}, and in accordance with the illustration described below Eq.~\eqref{eq:varphi4}, we obtain the four-particle collective phase from a linear combination of the triad phases. 

If there are two pairs of each mutually orthogonal particles, associated with two missing edges in the graph representation, two different cases can apply: Either the internal state of one particle is orthogonal to those of two other particles, say $\bracket{\phi_\alpha}{\phi_\gamma}=\bracket{\phi_\alpha}{\phi_\delta}=0$ [cf. Fig.~\ref{fig:COLP}(c)], or the particles are pairwise distininguishable, say $\bracket{\phi_\alpha}{\phi_\gamma}=\bracket{\phi_\beta}{\phi_\delta}=0$ [cf. Fig.~\ref{fig:COLP}(d)]. In both cases only a single pair of collective phases exists: a pair of triad phases $\varphi_{(\beta\gamma\delta)}$ and $\varphi_{(\beta\delta\gamma)}$ in the first, and four-particle collective phases $\varphi_{(\alpha\beta\gamma\delta)}$ and $\varphi_{(\alpha\delta\gamma\beta)}$ in the second case. Both pairs of collective phases correspond to inverse cycles, and, thus, are related by Eq.~\eqref{eq:varphi-}.

For graphs with three missing edges, only the connectivity illustrated in Fig.~\ref{fig:COLP}(e) gives rise to the presence of collective phases, namely a single pair of triad phases corresponding to inverse cycles. Again, according to Eq.~\eqref{eq:varphi-} these phases differ by their sign. Graphs with even more missing edges do not give rise to collective phases.

\subsection{Characterization of the collective phases and state reconstruction}\label{sec:pureRecon}
In the following we show how to extract all collective phases, apart from their sign, from the output statistics~\eqref{eq:pF} and~\eqref{eq:pA} obtained for input states with pure internal states. With the above relations between the collective phases, we then infer the signs of the extracted collective phases up to a global sign, and reconstruct the reduced external density operator $\rho_\mathrm{E}$ up to complex conjugation. 

For pure internal states $\rho_\alpha=\ketbra{\phi_\alpha}{\phi_\alpha}$ we have [see Eqs.~\eqref{eq:CollPhase} and~\eqref{eq:TrabyPure}, and recall that $\tr{\rho_\alpha \rho_\beta} \geq 0$]
\begin{align}\label{eq:TTTT}
T_{(\alpha \beta \cdots \gamma)} = \sqrt{T_{(\alpha\beta)}T_{(\beta\cdot)}\cdots T_{(\cdot\gamma)} T_{(\gamma\alpha)}},
\end{align}
with $T_{(\alpha\beta)}=\tr{\rho_\alpha \rho_\beta}=\abs{\bracket{\phi_\alpha}{\phi_\beta}}^2$. For the triad phase contained in $\braket{\Pi_\mathrm{B(F)}}_{(\alpha,\beta,\gamma)}$ from Eq.~\eqref{eq:W3p}, we then obtain
\begin{align}\label{eq:cosphich}
\cos\varphi_{(\alpha \beta \gamma)}=\frac{ 6\braket{\Pi_\mathrm{B(F)}}_{(\alpha,\beta,\gamma)} -1- T_{(\alpha\beta)}- T_{(\alpha\gamma)}- T_{(\beta\gamma)} }{ 2 \sqrt{T_{(\alpha\beta)}T_{(\beta\gamma)} T_{(\gamma\alpha)}} },
\end{align}
with the right hand side fully determined by the output statistics, according to Eqs.~\eqref{eq:Tabch},~\eqref{eq:W3pch}, and~\eqref{eq:W3pchF}. Since $\cos(\varphi)=\cos(-\varphi)$, Eq.~\eqref{eq:cosphich} yields (modulo $2\pi$) two solutions for the collective phase $\varphi_{(\alpha \beta \gamma)}$, which differ by their sign, 
\begin{align}\label{eq:triad16}
\varphi_{(\alpha \beta \gamma)}= \Phi_{(\alpha \beta \gamma)}\  \lor \ \varphi_{(\alpha \beta \gamma)}=-\Phi_{(\alpha \beta \gamma)},
\end{align}
with $\Phi_{(\alpha \beta \gamma)}\geq 0$. 

At this point, let us recall that the number of involved collective phases depends on the orthogonality between the particles' internal states, as illustrated by Fig.~\ref{fig:COLP}. In particular, by~\eqref{eq:CollPhase}, vanishing $T_{(\alpha\beta\cdots\gamma)}$ entails the absence of the collective phase $\varphi_{(\alpha\beta\cdots\gamma)}$. We now come back to our discussion in the previous section concerning all possible cases of reduced sets of collective phases due to such vanishing overlaps $T_{(\alpha\beta\cdots\gamma)}$.

First consider non-orthogonal internal states [see Fig.~\ref{fig:COLP}(a)]. In this case we have eight triad phases and six four-particle collective phases, corresponding to all cycles of length three and four, respectively. With the help of  Eq.~\eqref{eq:triad16} we then obtain $2^8$ different solutions (sign choices) for the set of all eight triad phases $\{\varphi_{(\alpha\beta\gamma)}\}_{(\alpha \beta \gamma)}$. Utilizing Eq.~\eqref{eq:varphi-} for triad phases with inverse particle labellings, the number of solutions reduces to $2^4=16$. Moreover, Eq.~\eqref{eq:varphi3} defines a consistency check for these $16$ solutions, such that only two solutions remain, 
\begin{align}\label{eq:varphichPM}
\{\varphi_{(\alpha\beta\gamma)}= \Phi_{(\alpha\beta\gamma)}^\mathrm{cc}\}_{(\alpha\beta\gamma)} \ \lor \ \{\varphi_{(\alpha\beta\gamma)}=- \Phi_{(\alpha\beta\gamma)}^\mathrm{cc} \}_{(\alpha\beta\gamma)},
\end{align}
with $\Phi_{(\alpha\beta\gamma)}^\mathrm{cc} \in \{ \Phi_{(\alpha\beta\gamma)},-\Phi_{(\alpha\beta\gamma)} \}$ the phases determined by the consistence check. With Eq.~\eqref{eq:varphi4} it hence follows that \emph{all} collective phases can be characterized, except for their global sign. In view of Eqs.~\eqref{eq:rhoElementTr} and~\eqref{eq:CollPhase}, together with Eqs.~\eqref{eq:Tabch} and~\eqref{eq:TTTT}, this allows us to fully reconstruct the reduced external density operator $\rho_\mathrm{E}$ from Eq.~\eqref{eq:rhoE}, up to complex conjugation, i.e., we obtain
\begin{align*}
\rho_\mathrm{E}\ \lor \ \rho_\mathrm{E}^*.
\end{align*}

Next consider the case where the internal states of particles $\alpha$ and $\gamma$ are orthogonal, i.e. $\bracket{\phi_\alpha}{\phi_\gamma}=0$ [see Fig.~\ref{fig:COLP}(b)]. As described in Sec.~\ref{sec:Pure-internal-states}, we have four triad phases and two four-particle collective phases [corresponding to all cycles of length three and four with a non-vanishing contribution~\eqref{eq:CollPhase}, respectively]. That is, Eq.~\eqref{eq:triad16} gives rise to $2^4$ solutions for the set of all four triad phases, which reduce to $2^2=4$ solutions with Eq.~\eqref{eq:varphi-}. On the other hand, much as in our derivation of Eq.~\eqref{eq:triad16}, we can start off from Eq.~\eqref{eq:T1234Statistics} and use Eqs.~\eqref{eq:Tabch} and~\eqref{eq:TTTT} to obtain two solutions for the four-particle collective phase,
\begin{align}\label{eq:ColPhase4ext}
\varphi_{(\beta\gamma\delta\alpha)}= \Phi_{(\beta\gamma\delta\alpha)}\  \lor \ \varphi_{(\beta\gamma\delta\alpha)}=-\Phi_{(\beta\gamma\delta\alpha)}.
\end{align}
Together with Eq.~\eqref{eq:ColPhasMisEdge} these two solutions allow us to perform a consistency check for the four solutions of the set of all four triad phases. Similar to the case of non-orthogonal internal states, this results in two solutions for the set of all four triad phases, which differ by their signs [similar to Eq.~\eqref{eq:varphichPM}]. That is, together with Eq.~\eqref{eq:ColPhasMisEdge} we can, again, reconstruct the full reduced external density operator $\rho_\mathrm{E}$, up to complex conjugation.

In all other cases with additional orthogonalities of the particles' internal states [see Figs.~\ref{fig:COLP}(c)-(e), and our discussion in Sec.~\ref{sec:Pure-internal-states}], there are at most two collective phases corresponding to inverse cycles. Accordingly, by Eq.~\eqref{eq:varphi-}, these phases only differ by their sign. Again, with Eq.~\eqref{eq:triad16} or~\eqref{eq:ColPhase4ext} we can characterize these phases up to their sign, and reconstruct the external density operator $\rho_\mathrm{E}$ up to complex conjugation.

Altogether, we can reconstruct the full reduced external density operator $\rho_\mathrm{E}$ from Eq.~\eqref{eq:rhoE}, up to complex conjugation, for any degree of distinguishability of the involved particles. As we show in Appendix~\ref{app:CC}, the highly symmetric structure of the hypercube unitary~\eqref{eq:U} entails an invariance of the transition probabilities $p_{\vec{S}}$ from Eq.~\eqref{eq:pS} under complex conjugation of $\rho_\mathrm{E}$. It is therefore not surprising that the associated output statistics does not distinguish $\rho_\mathrm{E}$ from its complex conjugate $\rho_\mathrm{E}^*$.

\section{Conclusion} \label{sec:Con}
The wide-ranging applicability of many-body indistinguishability for quantum information processing, together with the ever-increasing controllability of the particles' degrees of freedom, calls for versatile tools to certify and characterize their indistinguishability via interference patterns obtained from measurements which are blind to the particles' internal degrees of freedom which potentially make them distinguishable. However, many-body interference leads to intricate counting statistics, which aren't easily deciphered. Earlier approaches are therefore limited to the extraction of partial information on the particles' indistinguishability, leaving certain features such as collective phases unresolved. Here we overcame these limitations, and circumvented the complexity in the output statistics by the application of a highly symmetric unitary map. This produces well-structured output statistics with clear dependencies on the particles' indistinguishability. As a result we were able to fully characterize four-body indistinguishability, and eventually designed a method to reconstruct the many-body density operator up to complex conjugation. Since our characterization scheme builds on a non-interacting many-body transformation, it is suitable for diverse experimental platforms, e.g. for non-interacting ultracold atoms in optical lattices, or single photons in linear optics, and can readily be implemented in state-of-the-art linear optical interferometry equipped with photon number resolving detectors. 

Several questions remain for future research: Can our scheme be generalized for larger particle numbers? How robust is the extraction of the particles' indistinguishability under perturbations of the hypercube unitary? Are there any other unitary transformations which allow a full characterization of the particles' indistinguishability, and, if so, do these unitaries necessarily exhibit some symmetries? Instead of considering one particular unitary matrix, can we fully characterize the particles' indistinguishability by randomly sampling the unitary matrices, as suggested in \cite{Walschaers-SB-2016,Walschaers-FM-2016,Ketterer-CM-2019,Brydges-PR-2019}?

\begin{acknowledgements}
The authors thank Eric Brunner for fruitful discussions and Gabriel Dufour for helpful comments and proofreading the manuscript. C.D. acknowledges the Georg H. Endress foundation for financial support.
\end{acknowledgements}

\begin{appendix}

\section{Proof of Eqs.~(\ref{eq:Wa3p}) and~(\ref{eq:Wa2p})} \label{app:3p2p}
To prove Eqs.~\eqref{eq:Wa3p} and~\eqref{eq:Wa2p}, let us calculate the reduced external three- and two-particle state $\rho_\mathrm{E}^{3\mathrm{p}}$ and $\rho_\mathrm{E}^{2\mathrm{p}}$, respectively. By starting from the four-particle state $\rho_\mathrm{E}$ from Eq.~\eqref{eq:rhoE} and tracing out one particle, we find
\begin{align}
\rho_\mathrm{E}^{3\mathrm{p}}&=\trp{1}{\rho_\mathrm{E}} \nonumber \\
&=\sum_{\pi,\pi' \in \mathrm{S}_4} [\rho_\mathrm{E}]_{\pi,\pi'} \bracket{E_{\pi'(4)}}{E_{\pi(4)}}  \ketbra{\vec{E}_\pi^{3\mathrm{p}}}{\vec{E}_{\pi'}^{3\mathrm{p}}} \nonumber \\
&=\sum_{\substack{\pi,\pi' \in \mathrm{S}_4 \\ \pi(4)=\pi'(4)}} [\rho_\mathrm{E}]_{\pi,\pi'}  \ketbra{\vec{E}_\pi^{3\mathrm{p}}}{\vec{E}_{\pi'}^{3\mathrm{p}}}, \label{eq:rho3pE}
\end{align}
with $\ket{\vec{E}_\pi^{3\mathrm{p}}}=\ket{E_{\pi(1)},E_{\pi(2)},E_{\pi(3)}}$. Similarly, by tracing out one more particle, and using $\ket{\vec{E}_\pi^{2\mathrm{p}}}=\ket{E_{\pi(1)},E_{\pi(2)}}$, we obtain
\begin{align}
\rho_\mathrm{E}^{2\mathrm{p}}
&=\trp{2}{\rho_\mathrm{E}} \nonumber \\
&=\sum_{\substack{\pi,\pi' \in \mathrm{S}_4 \\ \pi(3)=\pi'(3) \\\pi(4)=\pi'(4)} } [\rho_\mathrm{E}]_{\pi,\pi'}  \ketbra{\vec{E}_\pi^{2\mathrm{p}}}{\vec{E}_{\pi'}^{2\mathrm{p}}}.\label{eq:rho2p}
\end{align}

Using Eq.~\eqref{eq:rho3pE} and the projection operator $\Pi_\mathrm{B(F)}=1/6 \sum_{\tau \in \mathrm{S}_3} (-1)^\tau_\mathrm{B(F)} \Pi_\tau$ onto the (anti)symmetric subspace of three particles, we obtain
\begin{align*}
&\phantom{=}\braket{\Pi_\mathrm{B(F)}}_{3\mathrm{p}}\\
&=\tr{\Pi_\mathrm{B(F)}\rho_\mathrm{E}^{3\mathrm{p}}} \\
&=\frac{1}{6}\sum_{\substack{\pi,\pi' \in \mathrm{S}_4 \\ \pi(4)=\pi'(4)}} \sum_{\tau \in \mathrm{S}_3} (-1)^\tau_\mathrm{B(F)} [\rho_\mathrm{E}]_{\pi,\pi'}  \bra{\vec{E}_{\pi'}^{3\mathrm{p}}}   \Pi_\tau  \ket{\vec{E}_{\pi}^{3\mathrm{p}}}.
\end{align*}
Since $\pi(4)=\pi'(4)$, the composition $\pi^{-1}\pi'$ belongs to the subgroup $\mathrm{S}_{\{1,2,3\}}\otimes \mathrm{S}_{\{4\}}$ of $\mathrm{S}_4$, which leaves $4$ invariant, and we have $\bra{\vec{E}_{\pi'}^{3\mathrm{p}}}   \Pi_\tau  \ket{\vec{E}_{\pi}^{3\mathrm{p}}} = \delta_{\tau,[\pi^{-1}\pi']_{123}}$, with $[\pi^{-1}\pi']_{123}$ referring to the first three elements of $\pi^{-1}\pi'$. Accordingly,
\begin{align}
\braket{\Pi_\mathrm{B(F)}}_{3\mathrm{p}}&=\frac{1}{6}\sum_{\substack{\pi,\pi' \in \mathrm{S}_4 \\ \pi(4)=\pi'(4)}}  (-1)^{\pi\pi'}_\mathrm{B(F)}\  [\rho_\mathrm{E}]_{\pi,\pi'} \nonumber \\
&=\frac{1}{6}\sum_{\alpha=1}^4\sum_{\substack{\pi,\pi' \in \mathrm{S}_4 \\ \pi(4)=\pi'(4)=\alpha}}  (-1)^{\pi(\pi')^{-1}}_\mathrm{B(F)} \ [\rho_\mathrm{E}]_{\pi(\pi')^{-1},\epsilon},\label{eq:W3pstep1}
\end{align}
where we used Eq.~\eqref{eq:rhoEPiKappa} in the last step. By $\pi(4)=\pi'(4)$, the composition $\pi(\pi')^{-1}$ belongs to the subgroup $\mathrm{S}_{\{1,2,3,4\}\setminus \alpha}\otimes \mathrm{S}_{\{\alpha\}}$ of $\mathrm{S}_4$, leaving $\alpha$ invariant. Hence, we can rewrite Eq.~\eqref{eq:W3pstep1} as
\begin{align*}
\braket{\Pi_\mathrm{B(F)}}_{3\mathrm{p}}&=\sum_{\alpha=1}^4\sum_{\kappa \in \mathrm{S}_{\{1,2,3,4\}\setminus \alpha}\otimes \mathrm{S}_{\{\alpha\}}}  (-1)^\kappa_\mathrm{B(F)} \ [\rho_\mathrm{E}]_{\kappa,\epsilon},
\end{align*}
and using Eqs.~\eqref{eq:rhoElementTr},~\eqref{eq:DL}, and~\eqref{eq:Dall}, we arrive at
\begin{align*}
\braket{\Pi_\mathrm{B(F)}}_\mathrm{3p}&=\frac{1}{6} \left[1+ \frac{1}{2} \ \mathcal{I}_{(1,1,2)} + \frac{1}{4} \ \mathcal{I}_{(1,3)} \right],
\end{align*}
which coincides with the sought-after relation from Eq.~\eqref{eq:Wa3p}. Similarly, one can show relation~\eqref{eq:Wa2p} for $\braket{\Pi_\mathrm{B(F)}}_\mathrm{2p}=\tr{\Pi_\mathrm{B(F)}\rho_\mathrm{E}^{2\mathrm{p}}} $ using the reduced external two-particle state~\eqref{eq:rho2p} and the projector $\Pi_\mathrm{B(F)}$ onto the (anti)symmetric subspace of two particles.

\section{Proof of Eq.~(\ref{eq:pSsigma-pS})} \label{app:pSsigma-pS}
To prove Eq.~\eqref{eq:pSsigma-pS}, let us consider the transition probabilities~\eqref{eq:pS} for the final output mode occupations $\vec{S}_{\sigma^{(\mathpzc{S})}}$, with $\sigma^{(\mathpzc{S})}$ from Eq.~\eqref{eq:sigma},
\begin{align}\label{eqapp:pSsigma} 
p_{\vec{S}_{\sigma^{(\mathpzc{S})}}}=S \sum_{\pi, \pi' \in \mathrm{S}_4} [\rho_\mathrm{E}]_{\pi,\pi'} \prod_{\alpha=1}^4 U_{ \sigma^{(\mathpzc{S})}(F_\alpha), \pi(\alpha)} U^*_{\sigma^{(\mathpzc{S})}(F_\alpha), \pi'(\alpha)}.
\end{align}
This expression differs from $p_{\vec{S}}$ in Eq.~\eqref{eq:pS} by the row-indices of the matrix elements of $U$. Note that the hypercube unitary $U$ remains invariant under transposition, i.e. $U_{j,k}=U_{k,j}$ [see Eqs.~\eqref{eq:U} and~\eqref{eq:Uelements}]. Moreover, let us consider its symmetry properties as presented in Eq.~(39) in \cite{Dittel-TD2-2018},
\begin{align*}
U_{k,\sigma^{(\mathpzc{S})}(j)}= U_{k,j}\  w(k,\mathpzc{S}) \exp(\mathrm{i}[\theta(\sigma^{(\mathpzc{S})}(j))-\theta(j)]),
\end{align*}
with $w(\cdot,\cdot)\in\{1,-1\}$ [see Eq.~\eqref{eq:Walsh} and Tab.~\eqref{tab:rad} below], and $\theta(\cdot)\in \mathbb{R}$ (note that the unitary in \cite{Dittel-TD2-2018} is defined as the transpose of $U$). Using 
\begin{align*}
\prod_{\gamma=1}^4 \exp(\mathrm{i}[\theta(\sigma^{(\mathpzc{S})}(F_\gamma))-\theta(F_\gamma)])=1,
\end{align*}
the product in~\eqref{eqapp:pSsigma} can be rewritten as
\begin{align}
&\prod_{\alpha=1}^4 U_{ \sigma^{(\mathpzc{S})}(F_\alpha), \pi(\alpha)} U^*_{\sigma^{(\mathpzc{S})}(F_\alpha), \pi'(\alpha)} \nonumber\\
\begin{split}
=& \prod_{\alpha=1}^4 U_{ F_\alpha, \pi(\alpha)} U^*_{F_\alpha, \pi'(\alpha)} \\
\times& \prod_{\beta=1}^4 w(\pi(\beta),\mathpzc{S}) w(\pi'(\beta),\mathpzc{S}). \label{eqapp:prodtwo}
\end{split}
\end{align}
Since $w(\cdot,\cdot)\in\{1,-1\}$, the second product in~\eqref{eqapp:prodtwo} yields 
\begin{align*}
\prod_{\beta=1}^4 w(\pi(\beta),\mathpzc{S}) w(\pi'(\beta),\mathpzc{S})&=\prod_{\beta=1}^4 w(\beta,\mathpzc{S}) w(\beta,\mathpzc{S})\\
&=1,
\end{align*}
such that
\begin{align}
\prod_{\alpha=1}^4 U_{ \sigma^{(\mathpzc{S})}(F_\alpha), \pi(\alpha)} U^*_{\sigma^{(\mathpzc{S})}(F_\alpha), \pi'(\alpha)} \nonumber
= \prod_{\alpha=1}^4 U_{ F_\alpha, \pi(\alpha)} U^*_{F_\alpha, \pi'(\alpha)}.
\end{align}
By plugging this into Eq.~\eqref{eqapp:pSsigma} and comparing with $p_{\vec{S}}$ from Eq.~\eqref{eq:pS}, we finally arrive at
\begin{align*}
p_{\vec{S}_{\sigma^{(\mathpzc{S})}}}&=S \sum_{\pi, \pi' \in \mathrm{S}_4} [\rho_\mathrm{E}]_{\pi,\pi'} \prod_{\alpha=1}^4 U_{ F_\alpha, \pi(\alpha)} U^*_{F_\alpha, \pi'(\alpha)}\\
&=p_{\vec{S}}.
\end{align*}
Given the Klein four-subgroup $\mathrm{K}_4=\{\epsilon, \sigma^{(1)},\sigma^{(2)},\sigma^{(3)}\}$ \cite{Baumslag-SO-1968} of $\mathrm{S}_4$, and, trivially, $ p_{\vec{S}_{\epsilon}}=p_{\vec{S}}$, this proves the sought-after relation from Eq.~\eqref{eq:pSsigma-pS}.

\section{Proof of the expressions in Eq.~(\ref{eq:pF})} \label{app:PF}
In the following, we prove the expressions~\eqref{eq:pF} for the transition probabilities of output events in the classes $\mathcal{F}_\mathpzc{S}^\mathrm{I(II)}$ from Eqs.~\eqref{eq:FclassI} and~\eqref{eq:FclassII}. First, we rewrite Eq.~\eqref{eq:pS},
\begin{align*}
p_{\vec{S}}&=S \sum_{\pi, \pi' \in \mathrm{S}_4} [\rho_\mathrm{E}]_{\pi,\pi'} \prod_{\alpha=1}^4 U_{F_\alpha, \pi(\alpha)} U^*_{F_\alpha, \pi'(\alpha)}\\
&=S \sum_{\pi, \pi' \in \mathrm{S}_4} [\rho_\mathrm{E}]_{\pi(\pi')^{-1},\epsilon} \prod_{\alpha=1}^4 U_{F_\alpha, \pi(\alpha)} U^*_{F_\alpha, \pi'(\alpha)},
\end{align*}
where $\epsilon$ is the identity permutation [see also Eq.~\eqref{eq:rhoEPiKappa}]. With $\kappa=\pi(\pi')^{-1}$, this becomes
\begin{align}
p_{\vec{S}}&=S \sum_{\kappa \in \mathrm{S}_4} [\rho_\mathrm{E}]_{\kappa,\epsilon} \sum_{\pi' \in \mathrm{S}_4}  \prod_{\alpha=1}^4 U_{F_\alpha, \kappa(\pi'(\alpha))} U^*_{F_\alpha, \pi'(\alpha)} \nonumber \\
&=\sum_{\kappa \in \mathrm{S}_4} [\rho_\mathrm{E}]_{\kappa,\epsilon} \  A(\kappa,\vec{S}),\label{eq:pSapp}
\end{align}
with 
\begin{align}\label{eq:Kappa}
A(\kappa,\vec{S})=S \sum_{\pi \in \mathrm{S}_4}  \prod_{\alpha=1}^4 U_{F_\alpha, \kappa(\pi(\alpha))} U^*_{F_\alpha, \pi(\alpha)}.
\end{align}

\begin{table}[t]
\begin{ruledtabular}
\begin{tabular}{l r r r r}
$j$ & $1$  & $2$  & $3$ & $4$  \\ \hline 
$w(j,1)=r(j,1)$ & $1$ & $1$ & $-1$ & $-1$  \\
$w(j,2)=r(j,2) $ & $1$ & $-1$ & $1$ & $-1$ \\
$w(j,3)=r(j,1)r(j,2)$ & $1$ & $-1$ & $-1$ & $1$ 
\end{tabular}
\caption{Rademacher functions $r(j,\mathpzc{S})$ and Walsh functions $w(j,\mathpzc{S})$ for all modes $j\in\{1,2,3,4\}$.}
\label{tab:rad}
\end{ruledtabular}
\end{table}

Next we utilize the elementwise expression of the hypercube unitary provided in \cite{Dittel-MB-2017} [see Eq.~(8) there],
\begin{align}\label{eq:Uelements}
U_{k,j}=\frac{1}{2} \exp\left( \mathrm{i} \frac{\pi}{4} \left[ 2-\sum_{\mathpzc{S}=1}^2 r(k,\mathpzc{S})r(j,\mathpzc{S})\right]\right).
\end{align}
Here $r(j,\mathpzc{S})$ denote the Rademacher functions \cite{Rademacher-ES-1922} defined for $\mathpzc{S}\in\{1,2\}$ as [see Eq.~(7) in \cite{Dittel-MB-2017}] 
\begin{align}\label{eq:rad}
r(j,\mathpzc{S})=(-1)^{\lfloor 2^\mathpzc{S}(j-1)/4 \rfloor  },
\end{align}
with the Gaussian brackets $\lfloor x\rfloor$ evaluating the greatest integer less than or equal to $x$. These functions are listed in Tab.~\ref{tab:rad} together with the Walsh functions \cite{Walsh-CS-1923} $w(j,\mathpzc{S})$, which are defined for all $j\in\{1,2,3,4\}$ and $\mathcal{S}\in\{1,2,3\}$ by [see Eq.~(12) in \cite{Dittel-MB-2017}]
\begin{align}\label{eq:Walsh}
\begin{split}
w(j,1)&=r(j,1)\\
w(j,2)&=r(j,2)\\
w(j,3)&=r(j,1)r(j,2).
\end{split}
\end{align}
Plugging Eq.~\eqref{eq:Uelements} into $A(\kappa,\vec{S})$ in Eq.~\eqref{eq:Kappa} then yields
\begin{align}\label{eq:KappaR}
\begin{split}
A(\kappa,\vec{S})&=\frac{S}{4^4} \sum_{\pi \in \mathrm{S}_4}  \exp\Bigg( \mathrm{i} \frac{\pi}{4} \sum_{\alpha=1}^4\sum_{\mathpzc{S}=1}^2 r(F_\alpha,\mathpzc{S}) \\
& \times \Big[ r(\pi(\alpha),\mathpzc{S}) - r( \kappa(\pi(\alpha)),\mathpzc{S} )    \Big] \Bigg).
\end{split}
\end{align}

\subsection{Probabilities for events contained in $\mathcal{F}_\mathpzc{S}^\mathrm{I}$}
Let us now  focus on the transition probabilities for output events in the class $\mathcal{F}_1^\mathrm{I}$. Since all of these output events appear with equal transition probability [recall Eq.~\eqref{eq:pSsigma-pS}], we take one representative of this class, $\vec{S}=(3,1,0,0)$, with corresponding mode assignment list $\vec{F}=(1,1,1,2)$. In this case,  we have $r(F_\alpha,1)=1$ for all $\alpha\in\{1,2,3,4\}$ [cf. Tab.~\ref{tab:rad}], such that for $\mathpzc{S}=1$ the sum over $\alpha$ in the exponent in~\eqref{eq:KappaR} vanishes, leaving only
\begin{align*}
\exp\left( \mathrm{i} \frac{\pi}{4} \sum_{\alpha=1}^4 r(F_\alpha,2)  \left[ r(\pi(\alpha),2) - r( \kappa(\pi(\alpha)),2 )    \right] \right).
\end{align*}
Moreover, we note that for $\mathpzc{S}=2$ we have $r(F_\alpha,2)=1$ for $\alpha\in\{1,2,3\}$, while $r(F_4,2)=r(2,2)=-1$. By expanding the sum in the exponent with 
\begin{align*}
0=\left[r(2,2)-r(2,2)\right]\left[r(\pi(4),2) - r( \kappa(\pi(4)),2 ) \right],
\end{align*}
the sum over $\alpha$ also vanishes for the symmetry $\mathpzc{S}=2$, however, we are left with
\begin{align*}
\exp\left( -\mathrm{i} \frac{\pi}{2}  \left[ r(\pi(4),2) - r( \kappa(\pi(4)),2 )    \right] \right).
\end{align*}
In view of $r(j,\mathpzc{S})\in \{1,-1\}$, this is equivalent to
\begin{align*}
r(\pi(4),2) \ r( \kappa(\pi(4)),2).
\end{align*}
With this simplification, and $S=4!/(3!1!0!0!)=4$, Eq.~\eqref{eq:KappaR} becomes
\begin{align}\label{eq:Kappa3100-1}
A(\kappa,(3,1,0,0))&=\frac{1}{4^3} \sum_{\pi \in \mathrm{S}_4} r(\pi(4),2) \ r( \kappa(\pi(4)),2 ).
\end{align}
Next, we note that for all $j\in\{1,2,3,4\}$, there are exactly $3!=6$ permutations in $\mathrm{S}_4$, for which $\pi(4)=j$. Hence, we can rewrite the sum in~\eqref{eq:Kappa3100-1} as a sum over all indices $j$,
\begin{align}\label{eq:Kappa3100-2}
A(\kappa,(3,1,0,0))&= \frac{6}{4^3}\sum_{j=1}^4 r(j,2) \ r( \kappa(j),2 ).
\end{align}

At this point, let us use the Walsh functions from Eq.~\eqref{eq:Walsh} and define
\begin{align}\label{eq:K}
K(\kappa,\mathpzc{S})=\sum_{j=1}^4 w(j,\mathpzc{S}) \ w( \kappa(j),\mathpzc{S} ),
\end{align}
which will be useful in the following. Using Eqs.~\eqref{eq:pSapp}, \eqref{eq:Kappa3100-2}, and~\eqref{eq:K}, the transition probability for output events in the class $\mathcal{F}_1^\mathrm{I}$ can be written as
\begin{align}\label{eq:pF1Iapp}
p_{\mathcal{F}_1^\mathrm{I}}=\frac{6}{4^3}  \sum_{\kappa \in \mathrm{S}_4} [\rho_\mathrm{E}]_{\kappa,\epsilon} \ K(\kappa,2).
\end{align} 
Here we see that the transition probability is readily obtained by evaluating $K(\kappa,\mathpzc{S})$ from Eq.~\eqref{eq:K}. This can be done with the help of Tab.~\ref{tab:rad}. To this end, let us define the sets of permutations
\begin{align*}
G_2^+&=\{\epsilon, (1~3), (2~4),(1~3)(2~4)\},\\
G_2^-&=\{ (1~2) (3~4),(1~4)(2~3),(1~4~3~2),(1~2~3~4)\},\\
G_2^0&=\mathrm{S}_4 \setminus (G_2^+ \cup G_2^- ),
\end{align*}
as well as (used further below)
\begin{align*}
G_1^+&=\{\epsilon, (1~2), (3~4),(1~2)(3~4)\},\\
G_1^-&=\{ (1~3) (2~4),(1~4)(2~3),(1~3~2~4),(1~4~2~3)\},\\
G_1^0&=\mathrm{S}_4 \setminus (G_1^+ \cup G_1^- ),
\end{align*}
and
\begin{align*}
G_3^+&=\{\epsilon, (1~4), (2~3),(1~4)(2~3)\},\\
G_3^-&=\{ (1~2) (3~4),(1~3)(2~4),(1~2~4~3),(1~3~4~2)\},\\
G_3^0&=\mathrm{S}_4 \setminus (G_3^+ \cup G_3^- ),
\end{align*}
for which we find
\begin{align} \label{eq:KappaPM4}
K(\kappa,\mathpzc{S})= \begin{cases} 
\phantom{-}4 \quad &\text{if} \quad \kappa \in G_\mathpzc{S}^+ \\
-4 \quad &\text{if} \quad \kappa \in G_\mathpzc{S}^- \\
\phantom{-}0 \quad &\text{if} \quad \kappa \in G_\mathpzc{S}^0.
  \end{cases}
\end{align}
Therewith, the transition probability~\eqref{eq:pF1Iapp} becomes
\begin{align*}
p_{\mathcal{F}_1^\mathrm{I}}&= \frac{6}{16} \left[ \sum_{\kappa \in G^+} [\rho_\mathrm{E}]_{\kappa,\epsilon} - \sum_{\kappa \in G^-} [\rho_\mathrm{E}]_{\kappa,\epsilon} \right]\\
&=\frac{1}{64} \Big[ 1\pm T_{(13)} \pm T_{(24)} + T_{(13)}T_{(24)} \\
&- T_{(12)}T_{(34)} -T_{(14)}T_{(23)}\mp  2 T_{(1234)} \cos\varphi_{(1234)} \Big],
\end{align*}
where we used Eqs.~\eqref{eq:rhoElementTr}, \eqref{eq:CollPhase}, and~\eqref{eq:trcos} in the last step. This coincides with the sought after relation of $p_{\mathcal{F}_1^\mathrm{I}}$ in~\eqref{eq:pF}. The expressions for $p_{\mathcal{F}_2^\mathrm{I}}$ and $p_{\mathcal{F}_3^\mathrm{I}}$ follow from similar considerations.

\subsection{Probabilities for events contained in $\mathcal{F}_\mathpzc{S}^\mathrm{II}$}
For output events in the classes $\mathcal{F}_\mathpzc{S}^\mathrm{II}$ we proceed similarly as above. Therefore, let us consider $\mathcal{F}_1^\mathrm{II}$, which contains the event $\vec{S}=(2,0,1,1)$ with corresponding mode assignment list $\vec{F}=(1,1,3,4)$. First, note that, for $\mathpzc{S}=2$, we have $r(F_\alpha,2)=1$ for $\alpha=\{1,2,3\}$, and $r(F_4,2)=r(4,2)=-1$. Thus, for the sum over $\alpha$ in the exponent in~\eqref{eq:KappaR} to vanish for $\mathpzc{S}=2$, we expand with
\begin{align*}
0=\left[ r(4,2)- r(4,2) \right] \left[ r(\pi(4),2)- r(\kappa(\pi(4)),2)  \right].
\end{align*}
Similarly, since for $\mathpzc{S}=1$ we have $r(F_\alpha,1)=1$ for $\alpha=\{1,2\}$, and $r(F_\alpha,1)=-1$ for $\alpha=\{3,4\}$, we additionally expand the sum in the exponent in~\eqref{eq:KappaR} with
\begin{align*}
0&=\left[ r(3,1)- r(3,1) \right] \left[ r(\pi(3),1)- r(\kappa(\pi(3)),1)  \right]\\
&+\left[ r(4,1)- r(4,1) \right] \left[ r(\pi(4),1)- r(\kappa(\pi(4)),1)  \right].
\end{align*}
Therewith, the exponent in~\eqref{eq:KappaR} becomes
\begin{align*}
\exp\Bigg(  -\mathrm{i} \frac{\pi}{2} \Big[ & \left[ r(\pi(4),2)- r(\kappa(\pi(4)),2)  \right] \\
+& \left[ r(\pi(3),1)- r(\kappa(\pi(3)),1)  \right] \\
+&  \left[ r(\pi(4),1)- r(\kappa(\pi(4)),1)  \right]   \Big] \Bigg).
\end{align*}
Again, using $r(j,\mathpzc{S})\in \{1,-1\}$, this is equivalent to
\begin{align*}
&  r(\pi(4),2)\  r(\kappa(\pi(4)),2) \ r(\pi(3),1)\ r(\kappa(\pi(3)),1)\\
\times & r(\pi(4),1)\ r(\kappa(\pi(4)),1)\\
=&r(\pi(3),1)\ r(\kappa(\pi(3)),1) \ w(\pi(4),3) \ w(\kappa(\pi(4)),3),
\end{align*}
where we used the definition~\eqref{eq:Walsh} of the Walsh functions in the last step. With this, and using $S=4!/(2!0!1!1!)=12$, we can write $A(\kappa,(2,0,1,1))$ from Eq.~\eqref{eq:KappaR} as
\begin{align*}
A(\kappa,\vec{S})=\frac{12}{4^4} \sum_{\pi \in \mathrm{S}_4}  & r(\pi(3),1)\ r(\kappa(\pi(3)),1) \\
\times & w(\pi(4),3) \ w(\kappa(\pi(4)),3).
\end{align*}
Here we note that for all $j,k\in\{1,2,3,4\}$, with $j\neq k$, the sum over $\mathrm{S}_4$ yields $2!=2$ times $\pi(3)=j$ and $\pi(4)=k$, such that
\begin{align*}
A(\kappa,\vec{S})&=\frac{24}{4^4} \sum_{\substack{j,k=1 \\j \neq k   }}^4   r(j,1)\ r(\kappa(j),1) \ w(k,3) \ w(\kappa(k),3)\\
&=\frac{3}{32} \Bigg[  \sum_{j,k=1}^4  r(j,1)\ r(\kappa(j),1) \ w(k,3) \ w(\kappa(k),3) \\
& \phantom{=\frac{3}{32} \Bigg[} -\sum_{l=1}^4   r(l,2)\ r(\kappa(l),2) \Bigg],
\end{align*}
where we used Eq.~\eqref{eq:Walsh} in the last step. With the definition of $K(\kappa,\mathpzc{S})$ from Eq.~\eqref{eq:K}, this can be written as
\begin{align*}
A(\kappa,\vec{S})&=\frac{3}{32} \left[ K(\kappa, 1)\ K(\kappa,3) - K(\kappa,2)\right],
\end{align*}
such that the transition probability~\eqref{eq:pSapp} for output events in the class $\mathcal{F}_1^\mathrm{II}$ becomes
\begin{align*}
p_{\mathcal{F}_1^\mathrm{II}}=\frac{3}{32} \sum_{\kappa \in \mathrm{S}_4} [\rho_\mathrm{E}]_{\kappa,\epsilon} \  \left[ K(\kappa, 1)\ K(\kappa,3) - K(\kappa,2)\right].
\end{align*}
Evaluating this expression with the help of Eq.~\eqref{eq:KappaPM4}, and using Eqs.~\eqref{eq:rhoElementTr}, \eqref{eq:CollPhase}, and~\eqref{eq:trcos}, finally leads to
\begin{align*}
p_{\mathcal{F}_1^\mathrm{II}}&=\frac{1}{64} \Big[3 \mp  T_{(13)} \mp T_{(24)} +3 T_{(13)}T_{(24)}\\
&-3 T_{(12)}T_{(34)} -3T_{(14)}T_{(23)}\pm 2T_{(1234)} \cos\varphi_{(1234)} \Big],
\end{align*}
which proves the expression for $p_{\mathcal{F}_1^\mathrm{II}}$ in~\eqref{eq:pF}. Again, the transition probabilities for output events in $p_{\mathcal{F}_2^\mathrm{II}}$ and $p_{\mathcal{F}_3^\mathrm{II}}$ follow from similar considerations.

\section{Invariance of the output statistics under complex conjugation of $\rho_\mathrm{E}$} \label{app:CC}
In the following, we show that the output statistics formed by the transition probabilities $p_{\vec{S}}$ from Eq.~\eqref{eq:pS} are invariant under complex conjugation of the reduced external density operator $\rho_\mathrm{E}$. Let us start with considering the elements $[\rho_\mathrm{E}]_{\kappa,\epsilon}$ from Eq.~\eqref{eq:rhoElementTr} and note that by Eq.~\eqref{eq:CollPhase} we have
\begin{align}\label{rhoEkappa-1}
[\rho_\mathrm{E}]_{\kappa,\epsilon}=[\rho_\mathrm{E}]_{\kappa^{-1},\epsilon}^*,
\end{align}
such that $[\rho_\mathrm{E}]_{\kappa,\epsilon} \in \mathbb{R}$ for self-inverse permutations $\kappa=\kappa^{-1}$. 

Next, we use the expression of the transition probabilities from Eq.~\eqref{eq:pSapp}, 
\begin{align}\label{eq:psA-1}
p_{\vec{S}}&=\sum_{\kappa \in \mathrm{S}_4} [\rho_\mathrm{E}]_{\kappa,\epsilon} \  A(\kappa,\vec{S}),
\end{align}
with $A(\kappa,\vec{S})$ from Eq.~\eqref{eq:Kappa}, satisfying
\begin{align}
A(\kappa,\vec{S})&=S \sum_{\pi \in \mathrm{S}_4}  \prod_{\alpha=1}^4 U_{F_\alpha, \kappa(\pi(\alpha))}\ U^*_{F_\alpha, \pi(\alpha)} \label{eq:AkappaS-11} \\
&=S\sum_{\pi' \in \mathrm{S}_4}  \prod_{\alpha=1}^4 U_{F_\alpha, \pi'(\alpha)}\ U^*_{F_\alpha, \kappa^{-1}(\pi'(\alpha))} \nonumber \\
&=A^*(\kappa^{-1},\vec{S}).\label{eq:Akappakappa-1}
\end{align}
Moreover, for the hypercube unitary~\eqref{eq:U} we find  $A(\kappa,\vec{S})\in\mathbb{R}$. In order to see this, let us write $U$ as provided in \cite{Dittel-TD2-2018},
\begin{align}\label{eq:UUS}
U=\Theta U^\mathrm{S} \Theta,
\end{align}
with $\Theta_{j,j}=\exp\left(\mathrm{i} \frac{\pi}{4} \sum_{l=1}^2[1-r(j,l)]\right)$, the Rademacher functions $r(\cdot,\cdot)$ from Eq.~\eqref{eq:rad}, and the Sylvester unitary
\begin{align}\label{eq:US}
U^\mathrm{S}=\frac{1}{2} \begin{pmatrix} 1&1\\1 & -1\end{pmatrix}^{\otimes 2}.
\end{align}
Plugging Eq.~\eqref{eq:UUS} into $A(\kappa,\vec{S})$ from Eq.~\eqref{eq:AkappaS-11} then leads to 
\begin{align*}
A(\kappa,\vec{S})&=S \sum_{\pi \in \mathrm{S}_4}  \prod_{\alpha=1}^4 U^\mathrm{S}_{F_\alpha, \kappa(\pi(\alpha))}\ (U^\mathrm{S}_{F_\alpha, \pi(\alpha)} )^*.
\end{align*}
Since $U^\mathrm{S}$ from Eq.~\eqref{eq:US} is a real matrix, we see that $A(\kappa,\vec{S})\in \mathbb{R}$.

In consideration of Eq.~\eqref{rhoEkappa-1} let us now split the sum in Eq.~\eqref{eq:psA-1} into a sum over self-inverse permutations $\kappa=\kappa^{-1}$, and a sum over pairs $\kappa,\kappa^{-1}$,
\begin{align*}
p_{\vec{S}}&=\sum_{\substack{\kappa \in \mathrm{S}_4 \\ \kappa=\kappa^{-1}}} [\rho_\mathrm{E}]_{\kappa,\epsilon} \  A(\kappa,\vec{S}) \\
&+ \sum_{\substack{\kappa \in \mathrm{S}_4 \\ \kappa<\kappa^{-1}}}   \left(A(\kappa,\vec{S}) [\rho_\mathrm{E}]_{\kappa,\epsilon} + A(\kappa^{-1},\vec{S}) [\rho_\mathrm{E}]_{\kappa^{-1},\epsilon} \right).
\end{align*}
Using Eqs.~\eqref{rhoEkappa-1} and~\eqref{eq:Akappakappa-1}, as well as $A(\kappa,\vec{S})\in \mathbb{R}$, this yields
\begin{align*}
p_{\vec{S}}&=\sum_{\substack{\kappa \in \mathrm{S}_4 \\ \kappa=\kappa^{-1}}} [\rho_\mathrm{E}]_{\kappa,\epsilon} \  A(\kappa,\vec{S}) \\
&+ \sum_{\substack{\kappa \in \mathrm{S}_4 \\ \kappa<\kappa^{-1}}}   A(\kappa,\vec{S})  \left( [\rho_\mathrm{E}]_{\kappa,\epsilon} +  [\rho_\mathrm{E}]_{\kappa,\epsilon}^* \right).
\end{align*}
Together with Eq.~\eqref{rhoEkappa-1}, this expression reveals that $p_{\vec{S}}$ remains invariant under complex conjugation of $\rho_\mathrm{E}$. Accordingly, the output statistics can contain no information from which we can distinguish between $\rho_\mathrm{E}$ and $\rho_\mathrm{E}^*$.

\end{appendix}


\begin{thebibliography}{55}%
\makeatletter
\providecommand \@ifxundefined [1]{%
 \@ifx{#1\undefined}
}%
\providecommand \@ifnum [1]{%
 \ifnum #1\expandafter \@firstoftwo
 \else \expandafter \@secondoftwo
 \fi
}%
\providecommand \@ifx [1]{%
 \ifx #1\expandafter \@firstoftwo
 \else \expandafter \@secondoftwo
 \fi
}%
\providecommand \natexlab [1]{#1}%
\providecommand \enquote  [1]{``#1''}%
\providecommand \bibnamefont  [1]{#1}%
\providecommand \bibfnamefont [1]{#1}%
\providecommand \citenamefont [1]{#1}%
\providecommand \href@noop [0]{\@secondoftwo}%
\providecommand \href [0]{\begingroup \@sanitize@url \@href}%
\providecommand \@href[1]{\@@startlink{#1}\@@href}%
\providecommand \@@href[1]{\endgroup#1\@@endlink}%
\providecommand \@sanitize@url [0]{\catcode `\\12\catcode `\$12\catcode
  `\&12\catcode `\#12\catcode `\^12\catcode `\_12\catcode `\%12\relax}%
\providecommand \@@startlink[1]{}%
\providecommand \@@endlink[0]{}%
\providecommand \url  [0]{\begingroup\@sanitize@url \@url }%
\providecommand \@url [1]{\endgroup\@href {#1}{\urlprefix }}%
\providecommand \urlprefix  [0]{URL }%
\providecommand \Eprint [0]{\href }%
\providecommand \doibase [0]{http://dx.doi.org/}%
\providecommand \selectlanguage [0]{\@gobble}%
\providecommand \bibinfo  [0]{\@secondoftwo}%
\providecommand \bibfield  [0]{\@secondoftwo}%
\providecommand \translation [1]{[#1]}%
\providecommand \BibitemOpen [0]{}%
\providecommand \bibitemStop [0]{}%
\providecommand \bibitemNoStop [0]{.\EOS\space}%
\providecommand \EOS [0]{\spacefactor3000\relax}%
\providecommand \BibitemShut  [1]{\csname bibitem#1\endcsname}%
\let\auto@bib@innerbib\@empty
\bibitem [{\citenamefont {Hong}, \citenamefont {Ou},\ and\ \citenamefont
  {Mandel}(1987)}]{Hong-MS-1987}%
  \BibitemOpen
  \bibfield  {author} {\bibinfo {author} {\bibfnamefont {C.~K.}\ \bibnamefont
  {Hong}}, \bibinfo {author} {\bibfnamefont {Z.~Y.}\ \bibnamefont {Ou}}, \ and\
  \bibinfo {author} {\bibfnamefont {L.}~\bibnamefont {Mandel}},\ }\href
  {\doibase 10.1103/PhysRevLett.59.2044} {\bibfield  {journal} {\bibinfo
  {journal} {Phys. Rev. Lett.}\ }\textbf {\bibinfo {volume} {59}},\ \bibinfo
  {pages} {2044} (\bibinfo {year} {1987})}\BibitemShut {NoStop}%
\bibitem [{\citenamefont {Shih}\ and\ \citenamefont
  {Alley}(1988)}]{Shih-NT-1988}%
  \BibitemOpen
  \bibfield  {author} {\bibinfo {author} {\bibfnamefont {Y.~H.}\ \bibnamefont
  {Shih}}\ and\ \bibinfo {author} {\bibfnamefont {C.~O.}\ \bibnamefont
  {Alley}},\ }\href {\doibase 10.1103/PhysRevLett.61.2921} {\bibfield
  {journal} {\bibinfo  {journal} {Phys. Rev. Lett.}\ }\textbf {\bibinfo
  {volume} {61}},\ \bibinfo {pages} {2921} (\bibinfo {year}
  {1988})}\BibitemShut {NoStop}%
\bibitem [{\citenamefont {Mayer}\ \emph {et~al.}(2011)\citenamefont {Mayer},
  \citenamefont {Tichy}, \citenamefont {Mintert}, \citenamefont {Konrad},\ and\
  \citenamefont {Buchleitner}}]{Mayer-CS-2011}%
  \BibitemOpen
  \bibfield  {author} {\bibinfo {author} {\bibfnamefont {K.}~\bibnamefont
  {Mayer}}, \bibinfo {author} {\bibfnamefont {M.~C.}\ \bibnamefont {Tichy}},
  \bibinfo {author} {\bibfnamefont {F.}~\bibnamefont {Mintert}}, \bibinfo
  {author} {\bibfnamefont {T.}~\bibnamefont {Konrad}}, \ and\ \bibinfo {author}
  {\bibfnamefont {A.}~\bibnamefont {Buchleitner}},\ }\href {\doibase
  10.1103/PhysRevA.83.062307} {\bibfield  {journal} {\bibinfo  {journal} {Phys.
  Rev. A}\ }\textbf {\bibinfo {volume} {83}},\ \bibinfo {pages} {062307}
  (\bibinfo {year} {2011})}\BibitemShut {NoStop}%
\bibitem [{\citenamefont {Bloch}, \citenamefont {Dalibard},\ and\ \citenamefont
  {Nascimb{\`e}ne}(2012)}]{Bloch-QS-2012}%
  \BibitemOpen
  \bibfield  {author} {\bibinfo {author} {\bibfnamefont {I.}~\bibnamefont
  {Bloch}}, \bibinfo {author} {\bibfnamefont {J.}~\bibnamefont {Dalibard}}, \
  and\ \bibinfo {author} {\bibfnamefont {S.}~\bibnamefont {Nascimb{\`e}ne}},\
  }\href {\doibase 10.1038/nphys2259} {\bibfield  {journal} {\bibinfo
  {journal} {Nat. Phys.}\ }\textbf {\bibinfo {volume} {8}},\ \bibinfo {pages}
  {267} (\bibinfo {year} {2012})}\BibitemShut {NoStop}%
\bibitem [{\citenamefont {Kaufman}\ \emph {et~al.}(2014)\citenamefont
  {Kaufman}, \citenamefont {Lester}, \citenamefont {Reynolds}, \citenamefont
  {Wall}, \citenamefont {Foss-Feig}, \citenamefont {Hazzard}, \citenamefont
  {Rey},\ and\ \citenamefont {Regal}}]{Kaufman-TW-2014}%
  \BibitemOpen
  \bibfield  {author} {\bibinfo {author} {\bibfnamefont {A.~M.}\ \bibnamefont
  {Kaufman}}, \bibinfo {author} {\bibfnamefont {B.~J.}\ \bibnamefont {Lester}},
  \bibinfo {author} {\bibfnamefont {C.~M.}\ \bibnamefont {Reynolds}}, \bibinfo
  {author} {\bibfnamefont {M.~L.}\ \bibnamefont {Wall}}, \bibinfo {author}
  {\bibfnamefont {M.}~\bibnamefont {Foss-Feig}}, \bibinfo {author}
  {\bibfnamefont {K.~R.~A.}\ \bibnamefont {Hazzard}}, \bibinfo {author}
  {\bibfnamefont {A.~M.}\ \bibnamefont {Rey}}, \ and\ \bibinfo {author}
  {\bibfnamefont {C.~A.}\ \bibnamefont {Regal}},\ }\href {\doibase
  10.1126/science.1250057} {\bibfield  {journal} {\bibinfo  {journal}
  {Science}\ }\textbf {\bibinfo {volume} {345}},\ \bibinfo {pages} {306}
  (\bibinfo {year} {2014})}\BibitemShut {NoStop}%
\bibitem [{\citenamefont {Preiss}\ \emph {et~al.}(2015)\citenamefont {Preiss},
  \citenamefont {Ma}, \citenamefont {Tai}, \citenamefont {Lukin}, \citenamefont
  {Rispoli}, \citenamefont {Zupancic}, \citenamefont {Lahini}, \citenamefont
  {Islam},\ and\ \citenamefont {Greiner}}]{Preiss-SC-2015}%
  \BibitemOpen
  \bibfield  {author} {\bibinfo {author} {\bibfnamefont {P.~M.}\ \bibnamefont
  {Preiss}}, \bibinfo {author} {\bibfnamefont {R.}~\bibnamefont {Ma}}, \bibinfo
  {author} {\bibfnamefont {M.~E.}\ \bibnamefont {Tai}}, \bibinfo {author}
  {\bibfnamefont {A.}~\bibnamefont {Lukin}}, \bibinfo {author} {\bibfnamefont
  {M.}~\bibnamefont {Rispoli}}, \bibinfo {author} {\bibfnamefont
  {P.}~\bibnamefont {Zupancic}}, \bibinfo {author} {\bibfnamefont
  {Y.}~\bibnamefont {Lahini}}, \bibinfo {author} {\bibfnamefont
  {R.}~\bibnamefont {Islam}}, \ and\ \bibinfo {author} {\bibfnamefont
  {M.}~\bibnamefont {Greiner}},\ }\href {\doibase 10.1126/science.1260364}
  {\bibfield  {journal} {\bibinfo  {journal} {Science}\ }\textbf {\bibinfo
  {volume} {347}},\ \bibinfo {pages} {1229} (\bibinfo {year}
  {2015})}\BibitemShut {NoStop}%
\bibitem [{\citenamefont {Islam}\ \emph {et~al.}(2015)\citenamefont {Islam},
  \citenamefont {Ma}, \citenamefont {Preiss}, \citenamefont {Tai},
  \citenamefont {Lukin}, \citenamefont {Rispoli},\ and\ \citenamefont
  {Greiner}}]{Islam-ME-2015}%
  \BibitemOpen
  \bibfield  {author} {\bibinfo {author} {\bibfnamefont {R.}~\bibnamefont
  {Islam}}, \bibinfo {author} {\bibfnamefont {R.}~\bibnamefont {Ma}}, \bibinfo
  {author} {\bibfnamefont {P.~M.}\ \bibnamefont {Preiss}}, \bibinfo {author}
  {\bibfnamefont {M.~E.}\ \bibnamefont {Tai}}, \bibinfo {author} {\bibfnamefont
  {A.}~\bibnamefont {Lukin}}, \bibinfo {author} {\bibfnamefont
  {M.}~\bibnamefont {Rispoli}}, \ and\ \bibinfo {author} {\bibfnamefont
  {M.}~\bibnamefont {Greiner}},\ }\href {\doibase 10.1038/nature15750}
  {\bibfield  {journal} {\bibinfo  {journal} {Nature}\ }\textbf {\bibinfo
  {volume} {528}},\ \bibinfo {pages} {77} (\bibinfo {year} {2015})}\BibitemShut
  {NoStop}%
\bibitem [{\citenamefont {Gross}\ and\ \citenamefont
  {Bloch}(2017)}]{Gross-QS-2017}%
  \BibitemOpen
  \bibfield  {author} {\bibinfo {author} {\bibfnamefont {C.}~\bibnamefont
  {Gross}}\ and\ \bibinfo {author} {\bibfnamefont {I.}~\bibnamefont {Bloch}},\
  }\href {\doibase 10.1126/science.aal3837} {\bibfield  {journal} {\bibinfo
  {journal} {Science}\ }\textbf {\bibinfo {volume} {357}},\ \bibinfo {pages}
  {995} (\bibinfo {year} {2017})}\BibitemShut {NoStop}%
\bibitem [{\citenamefont {Kaufman}\ \emph {et~al.}(2018)\citenamefont
  {Kaufman}, \citenamefont {Tichy}, \citenamefont {Mintert}, \citenamefont
  {Rey},\ and\ \citenamefont {Regal}}]{Kaufman-TH-2018}%
  \BibitemOpen
  \bibfield  {author} {\bibinfo {author} {\bibfnamefont {A.~M.}\ \bibnamefont
  {Kaufman}}, \bibinfo {author} {\bibfnamefont {M.~C.}\ \bibnamefont {Tichy}},
  \bibinfo {author} {\bibfnamefont {F.}~\bibnamefont {Mintert}}, \bibinfo
  {author} {\bibfnamefont {A.~M.}\ \bibnamefont {Rey}}, \ and\ \bibinfo
  {author} {\bibfnamefont {C.~A.}\ \bibnamefont {Regal}},\ }\href {\doibase
  https://doi.org/10.1016/bs.aamop.2018.03.003} {\ \bibinfo {series} {Adv.
  Atom. Mol. Opt. Phys.},\ \textbf {\bibinfo {volume} {67}},\ \bibinfo {pages}
  {377 } (\bibinfo {year} {2018})}\BibitemShut {NoStop}%
\bibitem [{\citenamefont {Preiss}\ \emph {et~al.}(2019)\citenamefont {Preiss},
  \citenamefont {Becher}, \citenamefont {Klemt}, \citenamefont {Klinkhamer},
  \citenamefont {Bergschneider}, \citenamefont {Defenu},\ and\ \citenamefont
  {Jochim}}]{Preiss-HC-2019}%
  \BibitemOpen
  \bibfield  {author} {\bibinfo {author} {\bibfnamefont {P.~M.}\ \bibnamefont
  {Preiss}}, \bibinfo {author} {\bibfnamefont {J.~H.}\ \bibnamefont {Becher}},
  \bibinfo {author} {\bibfnamefont {R.}~\bibnamefont {Klemt}}, \bibinfo
  {author} {\bibfnamefont {V.}~\bibnamefont {Klinkhamer}}, \bibinfo {author}
  {\bibfnamefont {A.}~\bibnamefont {Bergschneider}}, \bibinfo {author}
  {\bibfnamefont {N.}~\bibnamefont {Defenu}}, \ and\ \bibinfo {author}
  {\bibfnamefont {S.}~\bibnamefont {Jochim}},\ }\href {\doibase
  10.1103/PhysRevLett.122.143602} {\bibfield  {journal} {\bibinfo  {journal}
  {Phys. Rev. Lett.}\ }\textbf {\bibinfo {volume} {122}},\ \bibinfo {pages}
  {143602} (\bibinfo {year} {2019})}\BibitemShut {NoStop}%
\bibitem [{\citenamefont {Bergschneider}\ \emph {et~al.}(2019)\citenamefont
  {Bergschneider}, \citenamefont {Klinkhamer}, \citenamefont {Becher},
  \citenamefont {Klemt}, \citenamefont {Palm}, \citenamefont {Z{\"u}rn},
  \citenamefont {Jochim},\ and\ \citenamefont
  {Preiss}}]{Bergschneider-EC-2019}%
  \BibitemOpen
  \bibfield  {author} {\bibinfo {author} {\bibfnamefont {A.}~\bibnamefont
  {Bergschneider}}, \bibinfo {author} {\bibfnamefont {V.~M.}\ \bibnamefont
  {Klinkhamer}}, \bibinfo {author} {\bibfnamefont {J.~H.}\ \bibnamefont
  {Becher}}, \bibinfo {author} {\bibfnamefont {R.}~\bibnamefont {Klemt}},
  \bibinfo {author} {\bibfnamefont {L.}~\bibnamefont {Palm}}, \bibinfo {author}
  {\bibfnamefont {G.}~\bibnamefont {Z{\"u}rn}}, \bibinfo {author}
  {\bibfnamefont {S.}~\bibnamefont {Jochim}}, \ and\ \bibinfo {author}
  {\bibfnamefont {P.~M.}\ \bibnamefont {Preiss}},\ }\href {\doibase
  10.1038/s41567-019-0508-6} {\bibfield  {journal} {\bibinfo  {journal} {Nat.
  Phys.}\ }\textbf {\bibinfo {volume} {15}},\ \bibinfo {pages} {640} (\bibinfo
  {year} {2019})}\BibitemShut {NoStop}%
\bibitem [{\citenamefont {O'Brien}(2007)}]{OBrien-OQ-2007}%
  \BibitemOpen
  \bibfield  {author} {\bibinfo {author} {\bibfnamefont {J.~L.}\ \bibnamefont
  {O'Brien}},\ }\href {\doibase 10.1126/science.1142892} {\bibfield  {journal}
  {\bibinfo  {journal} {Science}\ }\textbf {\bibinfo {volume} {318}},\ \bibinfo
  {pages} {1567} (\bibinfo {year} {2007})}\BibitemShut {NoStop}%
\bibitem [{\citenamefont {O'Brien}, \citenamefont {Furusawa},\ and\
  \citenamefont {Vu{\v c}kovi{\'c}}(2009)}]{OBrien-PQ-2009}%
  \BibitemOpen
  \bibfield  {author} {\bibinfo {author} {\bibfnamefont {J.~L.}\ \bibnamefont
  {O'Brien}}, \bibinfo {author} {\bibfnamefont {A.}~\bibnamefont {Furusawa}}, \
  and\ \bibinfo {author} {\bibfnamefont {J.}~\bibnamefont {Vu{\v
  c}kovi{\'c}}},\ }\href {\doibase 10.1038/nphoton.2009.229} {\bibfield
  {journal} {\bibinfo  {journal} {Nat. Photonics}\ }\textbf {\bibinfo {volume}
  {3}},\ \bibinfo {pages} {687} (\bibinfo {year} {2009})}\BibitemShut {NoStop}%
\bibitem [{\citenamefont {Aaronson}\ and\ \citenamefont
  {Arkhipov}(2013)}]{Aaronson-CC-2013}%
  \BibitemOpen
  \bibfield  {author} {\bibinfo {author} {\bibfnamefont {S.}~\bibnamefont
  {Aaronson}}\ and\ \bibinfo {author} {\bibfnamefont {A.}~\bibnamefont
  {Arkhipov}},\ }\href {\doibase 10.4086/toc.2013.v009a004} {\bibfield
  {journal} {\bibinfo  {journal} {Theory of Comput.}\ }\textbf {\bibinfo
  {volume} {9}},\ \bibinfo {pages} {143} (\bibinfo {year} {2013})}\BibitemShut
  {NoStop}%
\bibitem [{\citenamefont {Carolan}\ \emph {et~al.}(2015)\citenamefont
  {Carolan}, \citenamefont {Harrold}, \citenamefont {Sparrow}, \citenamefont
  {Mart{\'\i}n-L{\'o}pez}, \citenamefont {Russell}, \citenamefont
  {Silverstone}, \citenamefont {Shadbolt}, \citenamefont {Matsudai},
  \citenamefont {Oguma}, \citenamefont {Itoh}, \citenamefont {Marshall},
  \citenamefont {Thompson}, \citenamefont {Matthews}, \citenamefont
  {Hashimoto}, \citenamefont {O{\textquoteright}Brien},\ and\ \citenamefont
  {Laing}}]{Carolan-UL-2015}%
  \BibitemOpen
  \bibfield  {author} {\bibinfo {author} {\bibfnamefont {J.}~\bibnamefont
  {Carolan}}, \bibinfo {author} {\bibfnamefont {C.}~\bibnamefont {Harrold}},
  \bibinfo {author} {\bibfnamefont {C.}~\bibnamefont {Sparrow}}, \bibinfo
  {author} {\bibfnamefont {E.}~\bibnamefont {Mart{\'\i}n-L{\'o}pez}}, \bibinfo
  {author} {\bibfnamefont {N.~J.}\ \bibnamefont {Russell}}, \bibinfo {author}
  {\bibfnamefont {J.~W.}\ \bibnamefont {Silverstone}}, \bibinfo {author}
  {\bibfnamefont {P.~J.}\ \bibnamefont {Shadbolt}}, \bibinfo {author}
  {\bibfnamefont {N.}~\bibnamefont {Matsudai}}, \bibinfo {author}
  {\bibfnamefont {M.}~\bibnamefont {Oguma}}, \bibinfo {author} {\bibfnamefont
  {M.}~\bibnamefont {Itoh}}, \bibinfo {author} {\bibfnamefont {G.~D.}\
  \bibnamefont {Marshall}}, \bibinfo {author} {\bibfnamefont {M.~G.}\
  \bibnamefont {Thompson}}, \bibinfo {author} {\bibfnamefont {J.~C.~F.}\
  \bibnamefont {Matthews}}, \bibinfo {author} {\bibfnamefont {T.}~\bibnamefont
  {Hashimoto}}, \bibinfo {author} {\bibfnamefont {J.~L.}\ \bibnamefont
  {O{\textquoteright}Brien}}, \ and\ \bibinfo {author} {\bibfnamefont
  {A.}~\bibnamefont {Laing}},\ }\href {\doibase 10.1126/science.aab3642}
  {\bibfield  {journal} {\bibinfo  {journal} {Science}\ }\textbf {\bibinfo
  {volume} {349}},\ \bibinfo {pages} {711} (\bibinfo {year}
  {2015})}\BibitemShut {NoStop}%
\bibitem [{\citenamefont {Flamini}, \citenamefont {Spagnolo},\ and\
  \citenamefont {Sciarrino}(2018)}]{Flamini-PQ-2018}%
  \BibitemOpen
  \bibfield  {author} {\bibinfo {author} {\bibfnamefont {F.}~\bibnamefont
  {Flamini}}, \bibinfo {author} {\bibfnamefont {N.}~\bibnamefont {Spagnolo}}, \
  and\ \bibinfo {author} {\bibfnamefont {F.}~\bibnamefont {Sciarrino}},\ }\href
  {\doibase 10.1088/1361-6633/aad5b2} {\bibfield  {journal} {\bibinfo
  {journal} {Rep. Prog. Phys.}\ }\textbf {\bibinfo {volume} {82}},\ \bibinfo
  {pages} {016001} (\bibinfo {year} {2018})}\BibitemShut {NoStop}%
\bibitem [{\citenamefont {Slussarenko}\ and\ \citenamefont
  {Pryde}(2019)}]{Slussarenko-PQ-2019}%
  \BibitemOpen
  \bibfield  {author} {\bibinfo {author} {\bibfnamefont {S.}~\bibnamefont
  {Slussarenko}}\ and\ \bibinfo {author} {\bibfnamefont {G.~J.}\ \bibnamefont
  {Pryde}},\ }\href {\doibase 10.1063/1.5115814} {\bibfield  {journal}
  {\bibinfo  {journal} {Appl. Phys. Rev.}\ }\textbf {\bibinfo {volume} {6}},\
  \bibinfo {pages} {041303} (\bibinfo {year} {2019})}\BibitemShut {NoStop}%
\bibitem [{\citenamefont {Tichy}(2015)}]{Tichy-SP-2015}%
  \BibitemOpen
  \bibfield  {author} {\bibinfo {author} {\bibfnamefont {M.~C.}\ \bibnamefont
  {Tichy}},\ }\href {\doibase 10.1103/PhysRevA.91.022316} {\bibfield  {journal}
  {\bibinfo  {journal} {Phys. Rev. A}\ }\textbf {\bibinfo {volume} {91}},\
  \bibinfo {pages} {022316} (\bibinfo {year} {2015})}\BibitemShut {NoStop}%
\bibitem [{\citenamefont
  {Shchesnovich}(2015{\natexlab{a}})}]{Shchesnovich-PI-2015}%
  \BibitemOpen
  \bibfield  {author} {\bibinfo {author} {\bibfnamefont {V.~S.}\ \bibnamefont
  {Shchesnovich}},\ }\href {\doibase 10.1103/PhysRevA.91.013844} {\bibfield
  {journal} {\bibinfo  {journal} {Phys. Rev. A}\ }\textbf {\bibinfo {volume}
  {91}},\ \bibinfo {pages} {013844} (\bibinfo {year}
  {2015}{\natexlab{a}})}\BibitemShut {NoStop}%
\bibitem [{\citenamefont {Tillmann}\ \emph {et~al.}(2015)\citenamefont
  {Tillmann}, \citenamefont {Tan}, \citenamefont {Stoeckl}, \citenamefont
  {Sanders}, \citenamefont {de~Guise}, \citenamefont {Heilmann}, \citenamefont
  {Nolte}, \citenamefont {Szameit},\ and\ \citenamefont
  {Walther}}]{Tillmann-GM-2015}%
  \BibitemOpen
  \bibfield  {author} {\bibinfo {author} {\bibfnamefont {M.}~\bibnamefont
  {Tillmann}}, \bibinfo {author} {\bibfnamefont {S.-H.}\ \bibnamefont {Tan}},
  \bibinfo {author} {\bibfnamefont {S.~E.}\ \bibnamefont {Stoeckl}}, \bibinfo
  {author} {\bibfnamefont {B.~C.}\ \bibnamefont {Sanders}}, \bibinfo {author}
  {\bibfnamefont {H.}~\bibnamefont {de~Guise}}, \bibinfo {author}
  {\bibfnamefont {R.}~\bibnamefont {Heilmann}}, \bibinfo {author}
  {\bibfnamefont {S.}~\bibnamefont {Nolte}}, \bibinfo {author} {\bibfnamefont
  {A.}~\bibnamefont {Szameit}}, \ and\ \bibinfo {author} {\bibfnamefont
  {P.}~\bibnamefont {Walther}},\ }\href {\doibase 10.1103/PhysRevX.5.041015}
  {\bibfield  {journal} {\bibinfo  {journal} {Phys. Rev. X}\ }\textbf {\bibinfo
  {volume} {5}},\ \bibinfo {pages} {041015} (\bibinfo {year}
  {2015})}\BibitemShut {NoStop}%
\bibitem [{\citenamefont {Walschaers}, \citenamefont {Kuipers},\ and\
  \citenamefont {Buchleitner}(2016)}]{Walschaers-FM-2016}%
  \BibitemOpen
  \bibfield  {author} {\bibinfo {author} {\bibfnamefont {M.}~\bibnamefont
  {Walschaers}}, \bibinfo {author} {\bibfnamefont {J.}~\bibnamefont {Kuipers}},
  \ and\ \bibinfo {author} {\bibfnamefont {A.}~\bibnamefont {Buchleitner}},\
  }\href {\doibase 10.1103/PhysRevA.94.020104} {\bibfield  {journal} {\bibinfo
  {journal} {Phys. Rev. A}\ }\textbf {\bibinfo {volume} {94}},\ \bibinfo
  {pages} {020104} (\bibinfo {year} {2016})}\BibitemShut {NoStop}%
\bibitem [{\citenamefont {Khalid}\ \emph {et~al.}(2018)\citenamefont {Khalid},
  \citenamefont {Spivak}, \citenamefont {Sanders},\ and\ \citenamefont
  {de~Guise}}]{Khalid-PS-2018}%
  \BibitemOpen
  \bibfield  {author} {\bibinfo {author} {\bibfnamefont {A.}~\bibnamefont
  {Khalid}}, \bibinfo {author} {\bibfnamefont {D.}~\bibnamefont {Spivak}},
  \bibinfo {author} {\bibfnamefont {B.~C.}\ \bibnamefont {Sanders}}, \ and\
  \bibinfo {author} {\bibfnamefont {H.}~\bibnamefont {de~Guise}},\ }\href
  {\doibase 10.1103/PhysRevA.97.063802} {\bibfield  {journal} {\bibinfo
  {journal} {Phys. Rev. A}\ }\textbf {\bibinfo {volume} {97}},\ \bibinfo
  {pages} {063802} (\bibinfo {year} {2018})}\BibitemShut {NoStop}%
\bibitem [{\citenamefont {Dittel}\ \emph {et~al.}(2019)\citenamefont {Dittel},
  \citenamefont {Dufour}, \citenamefont {Weihs},\ and\ \citenamefont
  {Buchleitner}}]{Dittel-WP-2018}%
  \BibitemOpen
  \bibfield  {author} {\bibinfo {author} {\bibfnamefont {C.}~\bibnamefont
  {Dittel}}, \bibinfo {author} {\bibfnamefont {G.}~\bibnamefont {Dufour}},
  \bibinfo {author} {\bibfnamefont {G.}~\bibnamefont {Weihs}}, \ and\ \bibinfo
  {author} {\bibfnamefont {A.}~\bibnamefont {Buchleitner}},\ }\href
  {https://arxiv.org/abs/1901.02810} {\bibfield  {journal} {\bibinfo  {journal}
  {arXiv:1901.02810}\ } (\bibinfo {year} {2019})}\BibitemShut {NoStop}%
\bibitem [{\citenamefont {Shchesnovich}\ and\ \citenamefont
  {Bezerra}(2018)}]{Shchesnovich-CP-2018}%
  \BibitemOpen
  \bibfield  {author} {\bibinfo {author} {\bibfnamefont {V.~S.}\ \bibnamefont
  {Shchesnovich}}\ and\ \bibinfo {author} {\bibfnamefont {M.~E.~O.}\
  \bibnamefont {Bezerra}},\ }\href {\doibase 10.1103/PhysRevA.98.033805}
  {\bibfield  {journal} {\bibinfo  {journal} {Phys. Rev. A}\ }\textbf {\bibinfo
  {volume} {98}},\ \bibinfo {pages} {033805} (\bibinfo {year}
  {2018})}\BibitemShut {NoStop}%
\bibitem [{\citenamefont {Walschaers}\ \emph {et~al.}(2016)\citenamefont
  {Walschaers}, \citenamefont {Kuipers}, \citenamefont {Urbina}, \citenamefont
  {Mayer}, \citenamefont {Tichy}, \citenamefont {Richter},\ and\ \citenamefont
  {Buchleitner}}]{Walschaers-SB-2016}%
  \BibitemOpen
  \bibfield  {author} {\bibinfo {author} {\bibfnamefont {M.}~\bibnamefont
  {Walschaers}}, \bibinfo {author} {\bibfnamefont {J.}~\bibnamefont {Kuipers}},
  \bibinfo {author} {\bibfnamefont {J.-D.}\ \bibnamefont {Urbina}}, \bibinfo
  {author} {\bibfnamefont {K.}~\bibnamefont {Mayer}}, \bibinfo {author}
  {\bibfnamefont {M.~C.}\ \bibnamefont {Tichy}}, \bibinfo {author}
  {\bibfnamefont {K.}~\bibnamefont {Richter}}, \ and\ \bibinfo {author}
  {\bibfnamefont {A.}~\bibnamefont {Buchleitner}},\ }\href
  {http://stacks.iop.org/1367-2630/18/i=3/a=032001} {\bibfield  {journal}
  {\bibinfo  {journal} {New J. Phys.}\ }\textbf {\bibinfo {volume} {18}},\
  \bibinfo {pages} {032001} (\bibinfo {year} {2016})}\BibitemShut {NoStop}%
\bibitem [{\citenamefont {Giordani}\ \emph {et~al.}(2018)\citenamefont
  {Giordani}, \citenamefont {Flamini}, \citenamefont {Pompili}, \citenamefont
  {Viggianiello}, \citenamefont {Spagnolo}, \citenamefont {Crespi},
  \citenamefont {Osellame}, \citenamefont {Wiebe}, \citenamefont {Walschaers},
  \citenamefont {Buchleitner},\ and\ \citenamefont
  {Sciarrino}}]{Giordani-ES-2018}%
  \BibitemOpen
  \bibfield  {author} {\bibinfo {author} {\bibfnamefont {T.}~\bibnamefont
  {Giordani}}, \bibinfo {author} {\bibfnamefont {F.}~\bibnamefont {Flamini}},
  \bibinfo {author} {\bibfnamefont {M.}~\bibnamefont {Pompili}}, \bibinfo
  {author} {\bibfnamefont {N.}~\bibnamefont {Viggianiello}}, \bibinfo {author}
  {\bibfnamefont {N.}~\bibnamefont {Spagnolo}}, \bibinfo {author}
  {\bibfnamefont {A.}~\bibnamefont {Crespi}}, \bibinfo {author} {\bibfnamefont
  {R.}~\bibnamefont {Osellame}}, \bibinfo {author} {\bibfnamefont
  {N.}~\bibnamefont {Wiebe}}, \bibinfo {author} {\bibfnamefont
  {M.}~\bibnamefont {Walschaers}}, \bibinfo {author} {\bibfnamefont
  {A.}~\bibnamefont {Buchleitner}}, \ and\ \bibinfo {author} {\bibfnamefont
  {F.}~\bibnamefont {Sciarrino}},\ }\href
  {https://doi.org/10.1038/s41566-018-0097-4} {\bibfield  {journal} {\bibinfo
  {journal} {Nat. Photonics}\ }\textbf {\bibinfo {volume} {12}},\ \bibinfo
  {pages} {173} (\bibinfo {year} {2018})}\BibitemShut {NoStop}%
\bibitem [{\citenamefont {Walschaers}(2020)}]{Walschaers-SM-2020}%
  \BibitemOpen
  \bibfield  {author} {\bibinfo {author} {\bibfnamefont {M.}~\bibnamefont
  {Walschaers}},\ }\href {\doibase 10.1088/1361-6455/ab5c30} {\bibfield
  {journal} {\bibinfo  {journal} {J. Phys. B: At., Mol. Opt. Phys.}\ }\textbf
  {\bibinfo {volume} {53}},\ \bibinfo {pages} {043001} (\bibinfo {year}
  {2020})}\BibitemShut {NoStop}%
\bibitem [{\citenamefont {{Brunner et al.}}(2021)}]{Brunner-CD-2021}%
  \BibitemOpen
  \bibfield  {author} {\bibinfo {author} {\bibfnamefont {E.}~\bibnamefont
  {{Brunner et al.}}},\ }\href@noop {} {\bibfield  {journal} {\bibinfo
  {journal} {in preparation}\ } (\bibinfo {year} {2021})}\BibitemShut {NoStop}%
\bibitem [{\citenamefont {Tichy}\ \emph {et~al.}(2010)\citenamefont {Tichy},
  \citenamefont {Tiersch}, \citenamefont {de~Melo}, \citenamefont {Mintert},\
  and\ \citenamefont {Buchleitner}}]{Tichy-ZT-2010}%
  \BibitemOpen
  \bibfield  {author} {\bibinfo {author} {\bibfnamefont {M.~C.}\ \bibnamefont
  {Tichy}}, \bibinfo {author} {\bibfnamefont {M.}~\bibnamefont {Tiersch}},
  \bibinfo {author} {\bibfnamefont {F.}~\bibnamefont {de~Melo}}, \bibinfo
  {author} {\bibfnamefont {F.}~\bibnamefont {Mintert}}, \ and\ \bibinfo
  {author} {\bibfnamefont {A.}~\bibnamefont {Buchleitner}},\ }\href {\doibase
  10.1103/PhysRevLett.104.220405} {\bibfield  {journal} {\bibinfo  {journal}
  {Phys. Rev. Lett.}\ }\textbf {\bibinfo {volume} {104}},\ \bibinfo {pages}
  {220405} (\bibinfo {year} {2010})}\BibitemShut {NoStop}%
\bibitem [{\citenamefont {Tichy}\ \emph {et~al.}(2012)\citenamefont {Tichy},
  \citenamefont {Tiersch}, \citenamefont {Mintert},\ and\ \citenamefont
  {Buchleitner}}]{Tichy-MP-2012}%
  \BibitemOpen
  \bibfield  {author} {\bibinfo {author} {\bibfnamefont {M.~C.}\ \bibnamefont
  {Tichy}}, \bibinfo {author} {\bibfnamefont {M.}~\bibnamefont {Tiersch}},
  \bibinfo {author} {\bibfnamefont {F.}~\bibnamefont {Mintert}}, \ and\
  \bibinfo {author} {\bibfnamefont {A.}~\bibnamefont {Buchleitner}},\ }\href
  {http://stacks.iop.org/1367-2630/14/i=9/a=093015} {\bibfield  {journal}
  {\bibinfo  {journal} {New J. Phys.}\ }\textbf {\bibinfo {volume} {14}},\
  \bibinfo {pages} {093015} (\bibinfo {year} {2012})}\BibitemShut {NoStop}%
\bibitem [{\citenamefont {Crespi}(2015)}]{Crespi-SL-2015}%
  \BibitemOpen
  \bibfield  {author} {\bibinfo {author} {\bibfnamefont {A.}~\bibnamefont
  {Crespi}},\ }\href {\doibase 10.1103/PhysRevA.91.013811} {\bibfield
  {journal} {\bibinfo  {journal} {Phys. Rev. A}\ }\textbf {\bibinfo {volume}
  {91}},\ \bibinfo {pages} {013811} (\bibinfo {year} {2015})}\BibitemShut
  {NoStop}%
\bibitem [{\citenamefont {Weimann}\ \emph {et~al.}(2016)\citenamefont
  {Weimann}, \citenamefont {Perez-Leija}, \citenamefont {Lebugle},
  \citenamefont {Keil}, \citenamefont {Tichy}, \citenamefont {Gr{\"a}fe},
  \citenamefont {Heilmann}, \citenamefont {Nolte}, \citenamefont {Moya-Cessa},
  \citenamefont {Weihs}, \citenamefont {Christodoulides},\ and\ \citenamefont
  {Szameit}}]{Weimann-IQ-2016}%
  \BibitemOpen
  \bibfield  {author} {\bibinfo {author} {\bibfnamefont {S.}~\bibnamefont
  {Weimann}}, \bibinfo {author} {\bibfnamefont {A.}~\bibnamefont
  {Perez-Leija}}, \bibinfo {author} {\bibfnamefont {M.}~\bibnamefont
  {Lebugle}}, \bibinfo {author} {\bibfnamefont {R.}~\bibnamefont {Keil}},
  \bibinfo {author} {\bibfnamefont {M.}~\bibnamefont {Tichy}}, \bibinfo
  {author} {\bibfnamefont {M.}~\bibnamefont {Gr{\"a}fe}}, \bibinfo {author}
  {\bibfnamefont {R.}~\bibnamefont {Heilmann}}, \bibinfo {author}
  {\bibfnamefont {S.}~\bibnamefont {Nolte}}, \bibinfo {author} {\bibfnamefont
  {H.}~\bibnamefont {Moya-Cessa}}, \bibinfo {author} {\bibfnamefont
  {G.}~\bibnamefont {Weihs}}, \bibinfo {author} {\bibfnamefont {D.~N.}\
  \bibnamefont {Christodoulides}}, \ and\ \bibinfo {author} {\bibfnamefont
  {A.}~\bibnamefont {Szameit}},\ }\href {https://doi.org/10.1038/ncomms11027}
  {\bibfield  {journal} {\bibinfo  {journal} {Nat. Commun.}\ }\textbf {\bibinfo
  {volume} {7}},\ \bibinfo {pages} {11027} (\bibinfo {year}
  {2016})}\BibitemShut {NoStop}%
\bibitem [{\citenamefont {Dittel}, \citenamefont {Keil},\ and\ \citenamefont
  {Weihs}(2017)}]{Dittel-MB-2017}%
  \BibitemOpen
  \bibfield  {author} {\bibinfo {author} {\bibfnamefont {C.}~\bibnamefont
  {Dittel}}, \bibinfo {author} {\bibfnamefont {R.}~\bibnamefont {Keil}}, \ and\
  \bibinfo {author} {\bibfnamefont {G.}~\bibnamefont {Weihs}},\ }\href
  {\doibase 10.1088/2058-9565/aa540c} {\bibfield  {journal} {\bibinfo
  {journal} {Quantum Sci. Technol.}\ }\textbf {\bibinfo {volume} {2}},\
  \bibinfo {pages} {015003} (\bibinfo {year} {2017})}\BibitemShut {NoStop}%
\bibitem [{\citenamefont {Viggianiello}\ \emph
  {et~al.}(2018{\natexlab{a}})\citenamefont {Viggianiello}, \citenamefont
  {Flamini}, \citenamefont {Innocenti}, \citenamefont {Cozzolino},
  \citenamefont {Bentivegna}, \citenamefont {Spagnolo}, \citenamefont {Crespi},
  \citenamefont {Brod}, \citenamefont {Galv{\={a}}o}, \citenamefont
  {Osellame},\ and\ \citenamefont {Sciarrino}}]{Viggianiello-EG-2018}%
  \BibitemOpen
  \bibfield  {author} {\bibinfo {author} {\bibfnamefont {N.}~\bibnamefont
  {Viggianiello}}, \bibinfo {author} {\bibfnamefont {F.}~\bibnamefont
  {Flamini}}, \bibinfo {author} {\bibfnamefont {L.}~\bibnamefont {Innocenti}},
  \bibinfo {author} {\bibfnamefont {D.}~\bibnamefont {Cozzolino}}, \bibinfo
  {author} {\bibfnamefont {M.}~\bibnamefont {Bentivegna}}, \bibinfo {author}
  {\bibfnamefont {N.}~\bibnamefont {Spagnolo}}, \bibinfo {author}
  {\bibfnamefont {A.}~\bibnamefont {Crespi}}, \bibinfo {author} {\bibfnamefont
  {D.~J.}\ \bibnamefont {Brod}}, \bibinfo {author} {\bibfnamefont {E.~F.}\
  \bibnamefont {Galv{\={a}}o}}, \bibinfo {author} {\bibfnamefont
  {R.}~\bibnamefont {Osellame}}, \ and\ \bibinfo {author} {\bibfnamefont
  {F.}~\bibnamefont {Sciarrino}},\ }\href
  {http://stacks.iop.org/1367-2630/20/i=3/a=033017} {\bibfield  {journal}
  {\bibinfo  {journal} {New J. Phys.}\ }\textbf {\bibinfo {volume} {20}},\
  \bibinfo {pages} {033017} (\bibinfo {year} {2018}{\natexlab{a}})}\BibitemShut
  {NoStop}%
\bibitem [{\citenamefont {Tschernig}\ \emph {et~al.}(2018)\citenamefont
  {Tschernig}, \citenamefont {de~J.~Le\'{o}n-Montiel}, \citenamefont {{n}a
  Loaiza}, \citenamefont {Szameit}, \citenamefont {Busch},\ and\ \citenamefont
  {Perez-Leija}}]{Tschernig-MD-2018}%
  \BibitemOpen
  \bibfield  {author} {\bibinfo {author} {\bibfnamefont {K.}~\bibnamefont
  {Tschernig}}, \bibinfo {author} {\bibfnamefont {R.}~\bibnamefont
  {de~J.~Le\'{o}n-Montiel}}, \bibinfo {author} {\bibfnamefont {O.~S.~M.}\
  \bibnamefont {{n}a Loaiza}}, \bibinfo {author} {\bibfnamefont
  {A.}~\bibnamefont {Szameit}}, \bibinfo {author} {\bibfnamefont
  {K.}~\bibnamefont {Busch}}, \ and\ \bibinfo {author} {\bibfnamefont
  {A.}~\bibnamefont {Perez-Leija}},\ }\href {\doibase 10.1364/JOSAB.35.001985}
  {\bibfield  {journal} {\bibinfo  {journal} {J. Opt. Soc. Am. B}\ }\textbf
  {\bibinfo {volume} {35}},\ \bibinfo {pages} {1985} (\bibinfo {year}
  {2018})}\BibitemShut {NoStop}%
\bibitem [{\citenamefont {Dittel}\ \emph
  {et~al.}(2018{\natexlab{a}})\citenamefont {Dittel}, \citenamefont {Dufour},
  \citenamefont {Walschaers}, \citenamefont {Weihs}, \citenamefont
  {Buchleitner},\ and\ \citenamefont {Keil}}]{Dittel-TD1-2018}%
  \BibitemOpen
  \bibfield  {author} {\bibinfo {author} {\bibfnamefont {C.}~\bibnamefont
  {Dittel}}, \bibinfo {author} {\bibfnamefont {G.}~\bibnamefont {Dufour}},
  \bibinfo {author} {\bibfnamefont {M.}~\bibnamefont {Walschaers}}, \bibinfo
  {author} {\bibfnamefont {G.}~\bibnamefont {Weihs}}, \bibinfo {author}
  {\bibfnamefont {A.}~\bibnamefont {Buchleitner}}, \ and\ \bibinfo {author}
  {\bibfnamefont {R.}~\bibnamefont {Keil}},\ }\href {\doibase
  10.1103/PhysRevLett.120.240404} {\bibfield  {journal} {\bibinfo  {journal}
  {Phys. Rev. Lett.}\ }\textbf {\bibinfo {volume} {120}},\ \bibinfo {pages}
  {240404} (\bibinfo {year} {2018}{\natexlab{a}})}\BibitemShut {NoStop}%
\bibitem [{\citenamefont {Dittel}\ \emph
  {et~al.}(2018{\natexlab{b}})\citenamefont {Dittel}, \citenamefont {Dufour},
  \citenamefont {Walschaers}, \citenamefont {Weihs}, \citenamefont
  {Buchleitner},\ and\ \citenamefont {Keil}}]{Dittel-TD2-2018}%
  \BibitemOpen
  \bibfield  {author} {\bibinfo {author} {\bibfnamefont {C.}~\bibnamefont
  {Dittel}}, \bibinfo {author} {\bibfnamefont {G.}~\bibnamefont {Dufour}},
  \bibinfo {author} {\bibfnamefont {M.}~\bibnamefont {Walschaers}}, \bibinfo
  {author} {\bibfnamefont {G.}~\bibnamefont {Weihs}}, \bibinfo {author}
  {\bibfnamefont {A.}~\bibnamefont {Buchleitner}}, \ and\ \bibinfo {author}
  {\bibfnamefont {R.}~\bibnamefont {Keil}},\ }\href {\doibase
  10.1103/PhysRevA.97.062116} {\bibfield  {journal} {\bibinfo  {journal} {Phys.
  Rev. A}\ }\textbf {\bibinfo {volume} {97}},\ \bibinfo {pages} {062116}
  (\bibinfo {year} {2018}{\natexlab{b}})}\BibitemShut {NoStop}%
\bibitem [{\citenamefont {Ehrhardt}\ \emph {et~al.}(2021)\citenamefont
  {Ehrhardt}, \citenamefont {Keil}, \citenamefont {Maczewsky}, \citenamefont
  {Dittel}, \citenamefont {Heinrich},\ and\ \citenamefont
  {Szameit}}]{Ehrhardt-EC-2020}%
  \BibitemOpen
  \bibfield  {author} {\bibinfo {author} {\bibfnamefont {M.}~\bibnamefont
  {Ehrhardt}}, \bibinfo {author} {\bibfnamefont {R.}~\bibnamefont {Keil}},
  \bibinfo {author} {\bibfnamefont {L.}~\bibnamefont {Maczewsky}}, \bibinfo
  {author} {\bibfnamefont {C.}~\bibnamefont {Dittel}}, \bibinfo {author}
  {\bibfnamefont {M.}~\bibnamefont {Heinrich}}, \ and\ \bibinfo {author}
  {\bibfnamefont {A.}~\bibnamefont {Szameit}},\ }\href {\doibase
  10.1126/sciadv.abc5266} {\bibfield  {journal} {\bibinfo  {journal} {Science
  Advances}\ }\textbf {\bibinfo {volume} {7}},\ \bibinfo {pages} {eabc5266}
  (\bibinfo {year} {2021})}\BibitemShut {NoStop}%
\bibitem [{\citenamefont {M\"unzberg}\ \emph {et~al.}(2021)\citenamefont
  {M\"unzberg}, \citenamefont {Dittel}, \citenamefont {Lebugle}, \citenamefont
  {Buchleitner}, \citenamefont {Szameit}, \citenamefont {Weihs},\ and\
  \citenamefont {Keil}}]{Muenzberg-SI-2021}%
  \BibitemOpen
  \bibfield  {author} {\bibinfo {author} {\bibfnamefont {J.}~\bibnamefont
  {M\"unzberg}}, \bibinfo {author} {\bibfnamefont {C.}~\bibnamefont {Dittel}},
  \bibinfo {author} {\bibfnamefont {M.}~\bibnamefont {Lebugle}}, \bibinfo
  {author} {\bibfnamefont {A.}~\bibnamefont {Buchleitner}}, \bibinfo {author}
  {\bibfnamefont {A.}~\bibnamefont {Szameit}}, \bibinfo {author} {\bibfnamefont
  {G.}~\bibnamefont {Weihs}}, \ and\ \bibinfo {author} {\bibfnamefont
  {R.}~\bibnamefont {Keil}},\ }\href {\doibase 10.1103/PRXQuantum.2.020326}
  {\bibfield  {journal} {\bibinfo  {journal} {PRX Quantum}\ }\textbf {\bibinfo
  {volume} {2}},\ \bibinfo {pages} {020326} (\bibinfo {year}
  {2021})}\BibitemShut {NoStop}%
\bibitem [{\citenamefont {Brod}\ \emph {et~al.}(2019)\citenamefont {Brod},
  \citenamefont {Galv{\~a}o}, \citenamefont {Viggianiello}, \citenamefont
  {Flamini}, \citenamefont {Spagnolo},\ and\ \citenamefont
  {Sciarrino}}]{Brod-WG-2019}%
  \BibitemOpen
  \bibfield  {author} {\bibinfo {author} {\bibfnamefont {D.~J.}\ \bibnamefont
  {Brod}}, \bibinfo {author} {\bibfnamefont {E.~F.}\ \bibnamefont
  {Galv{\~a}o}}, \bibinfo {author} {\bibfnamefont {N.}~\bibnamefont
  {Viggianiello}}, \bibinfo {author} {\bibfnamefont {F.}~\bibnamefont
  {Flamini}}, \bibinfo {author} {\bibfnamefont {N.}~\bibnamefont {Spagnolo}}, \
  and\ \bibinfo {author} {\bibfnamefont {F.}~\bibnamefont {Sciarrino}},\ }\href
  {\doibase 10.1103/PhysRevLett.122.063602} {\bibfield  {journal} {\bibinfo
  {journal} {Phys. Rev. Lett.}\ }\textbf {\bibinfo {volume} {122}},\ \bibinfo
  {pages} {063602} (\bibinfo {year} {2019})}\BibitemShut {NoStop}%
\bibitem [{\citenamefont {Giordani}\ \emph {et~al.}(2020)\citenamefont
  {Giordani}, \citenamefont {Brod}, \citenamefont {Esposito}, \citenamefont
  {Viggianiello}, \citenamefont {Romano}, \citenamefont {Flamini},
  \citenamefont {Carvacho}, \citenamefont {Spagnolo}, \citenamefont
  {Galv{\~{a}}o},\ and\ \citenamefont {Sciarrino}}]{Giordani-EQ-2020}%
  \BibitemOpen
  \bibfield  {author} {\bibinfo {author} {\bibfnamefont {T.}~\bibnamefont
  {Giordani}}, \bibinfo {author} {\bibfnamefont {D.~J.}\ \bibnamefont {Brod}},
  \bibinfo {author} {\bibfnamefont {C.}~\bibnamefont {Esposito}}, \bibinfo
  {author} {\bibfnamefont {N.}~\bibnamefont {Viggianiello}}, \bibinfo {author}
  {\bibfnamefont {M.}~\bibnamefont {Romano}}, \bibinfo {author} {\bibfnamefont
  {F.}~\bibnamefont {Flamini}}, \bibinfo {author} {\bibfnamefont
  {G.}~\bibnamefont {Carvacho}}, \bibinfo {author} {\bibfnamefont
  {N.}~\bibnamefont {Spagnolo}}, \bibinfo {author} {\bibfnamefont {E.~F.}\
  \bibnamefont {Galv{\~{a}}o}}, \ and\ \bibinfo {author} {\bibfnamefont
  {F.}~\bibnamefont {Sciarrino}},\ }\href {\doibase 10.1088/1367-2630/ab7a30}
  {\bibfield  {journal} {\bibinfo  {journal} {New J. Phys.}\ }\textbf {\bibinfo
  {volume} {22}},\ \bibinfo {pages} {043001} (\bibinfo {year}
  {2020})}\BibitemShut {NoStop}%
\bibitem [{\citenamefont
  {Shchesnovich}(2015{\natexlab{b}})}]{Shchesnovich-TB-2015}%
  \BibitemOpen
  \bibfield  {author} {\bibinfo {author} {\bibfnamefont {V.~S.}\ \bibnamefont
  {Shchesnovich}},\ }\href {\doibase 10.1103/PhysRevA.91.063842} {\bibfield
  {journal} {\bibinfo  {journal} {Phys. Rev. A}\ }\textbf {\bibinfo {volume}
  {91}},\ \bibinfo {pages} {063842} (\bibinfo {year}
  {2015}{\natexlab{b}})}\BibitemShut {NoStop}%
\bibitem [{\citenamefont {Stanisic}\ and\ \citenamefont
  {Turner}(2018)}]{Stanisic-DD-2018}%
  \BibitemOpen
  \bibfield  {author} {\bibinfo {author} {\bibfnamefont {S.}~\bibnamefont
  {Stanisic}}\ and\ \bibinfo {author} {\bibfnamefont {P.~S.}\ \bibnamefont
  {Turner}},\ }\href {\doibase 10.1103/PhysRevA.98.043839} {\bibfield
  {journal} {\bibinfo  {journal} {Phys. Rev. A}\ }\textbf {\bibinfo {volume}
  {98}},\ \bibinfo {pages} {043839} (\bibinfo {year} {2018})}\BibitemShut
  {NoStop}%
\bibitem [{\citenamefont {Viggianiello}\ \emph
  {et~al.}(2018{\natexlab{b}})\citenamefont {Viggianiello}, \citenamefont
  {Flamini}, \citenamefont {Bentivegna}, \citenamefont {Spagnolo},
  \citenamefont {Crespi}, \citenamefont {Brod}, \citenamefont {Galvão},
  \citenamefont {Osellame},\ and\ \citenamefont
  {Sciarrino}}]{Viggianiello-OP-2018}%
  \BibitemOpen
  \bibfield  {author} {\bibinfo {author} {\bibfnamefont {N.}~\bibnamefont
  {Viggianiello}}, \bibinfo {author} {\bibfnamefont {F.}~\bibnamefont
  {Flamini}}, \bibinfo {author} {\bibfnamefont {M.}~\bibnamefont {Bentivegna}},
  \bibinfo {author} {\bibfnamefont {N.}~\bibnamefont {Spagnolo}}, \bibinfo
  {author} {\bibfnamefont {A.}~\bibnamefont {Crespi}}, \bibinfo {author}
  {\bibfnamefont {D.~J.}\ \bibnamefont {Brod}}, \bibinfo {author}
  {\bibfnamefont {E.~F.}\ \bibnamefont {Galvão}}, \bibinfo {author}
  {\bibfnamefont {R.}~\bibnamefont {Osellame}}, \ and\ \bibinfo {author}
  {\bibfnamefont {F.}~\bibnamefont {Sciarrino}},\ }\href {\doibase
  https://doi.org/10.1016/j.scib.2018.10.009} {\bibfield  {journal} {\bibinfo
  {journal} {Science Bulletin}\ }\textbf {\bibinfo {volume} {63}},\ \bibinfo
  {pages} {1470 } (\bibinfo {year} {2018}{\natexlab{b}})}\BibitemShut {NoStop}%
\bibitem [{\citenamefont {Menssen}\ \emph {et~al.}(2017)\citenamefont
  {Menssen}, \citenamefont {Jones}, \citenamefont {Metcalf}, \citenamefont
  {Tichy}, \citenamefont {Barz}, \citenamefont {Kolthammer},\ and\
  \citenamefont {Walmsley}}]{Menssen-DM-2017}%
  \BibitemOpen
  \bibfield  {author} {\bibinfo {author} {\bibfnamefont {A.~J.}\ \bibnamefont
  {Menssen}}, \bibinfo {author} {\bibfnamefont {A.~E.}\ \bibnamefont {Jones}},
  \bibinfo {author} {\bibfnamefont {B.~J.}\ \bibnamefont {Metcalf}}, \bibinfo
  {author} {\bibfnamefont {M.~C.}\ \bibnamefont {Tichy}}, \bibinfo {author}
  {\bibfnamefont {S.}~\bibnamefont {Barz}}, \bibinfo {author} {\bibfnamefont
  {W.~S.}\ \bibnamefont {Kolthammer}}, \ and\ \bibinfo {author} {\bibfnamefont
  {I.~A.}\ \bibnamefont {Walmsley}},\ }\href {\doibase
  10.1103/PhysRevLett.118.153603} {\bibfield  {journal} {\bibinfo  {journal}
  {Phys. Rev. Lett.}\ }\textbf {\bibinfo {volume} {118}},\ \bibinfo {pages}
  {153603} (\bibinfo {year} {2017})}\BibitemShut {NoStop}%
\bibitem [{\citenamefont {Jones}\ \emph {et~al.}(2020)\citenamefont {Jones},
  \citenamefont {Menssen}, \citenamefont {Chrzanowski}, \citenamefont
  {Wolterink}, \citenamefont {Shchesnovich},\ and\ \citenamefont
  {Walmsley}}]{Jones-ID-2020}%
  \BibitemOpen
  \bibfield  {author} {\bibinfo {author} {\bibfnamefont {A.~E.}\ \bibnamefont
  {Jones}}, \bibinfo {author} {\bibfnamefont {A.~J.}\ \bibnamefont {Menssen}},
  \bibinfo {author} {\bibfnamefont {H.~M.}\ \bibnamefont {Chrzanowski}},
  \bibinfo {author} {\bibfnamefont {T.~A.~W.}\ \bibnamefont {Wolterink}},
  \bibinfo {author} {\bibfnamefont {V.~S.}\ \bibnamefont {Shchesnovich}}, \
  and\ \bibinfo {author} {\bibfnamefont {I.~A.}\ \bibnamefont {Walmsley}},\
  }\href {\doibase 10.1103/PhysRevLett.125.123603} {\bibfield  {journal}
  {\bibinfo  {journal} {Phys. Rev. Lett.}\ }\textbf {\bibinfo {volume} {125}},\
  \bibinfo {pages} {123603} (\bibinfo {year} {2020})}\BibitemShut {NoStop}%
\bibitem [{\citenamefont {Dittel}(2019)}]{Dittel-AI-2019}%
  \BibitemOpen
  \bibfield  {author} {\bibinfo {author} {\bibfnamefont {C.}~\bibnamefont
  {Dittel}},\ }\emph {\bibinfo {title} {About the interference of many
  particles}},\ \href {https://resolver.obvsg.at/urn:nbn:at:at-ubi:1-47210}
  {Ph.D. thesis},\ \bibinfo  {school} {University of Innsbruck,
  urn:nbn:at:at-ubi:1-47210} (\bibinfo {year} {2019})\BibitemShut {NoStop}%
\bibitem [{\citenamefont {Dummit}\ and\ \citenamefont
  {Foote}(2003)}]{Dummit-AA-2003}%
  \BibitemOpen
  \bibfield  {author} {\bibinfo {author} {\bibfnamefont {D.~S.}\ \bibnamefont
  {Dummit}}\ and\ \bibinfo {author} {\bibfnamefont {R.~M.}\ \bibnamefont
  {Foote}},\ }\href@noop {} {\emph {\bibinfo {title} {Abstract Algebra}}}\
  (\bibinfo  {publisher} {Wiley},\ \bibinfo {year} {2003})\BibitemShut
  {NoStop}%
\bibitem [{Note1()}]{Note1}%
  \BibitemOpen
  \bibinfo {note} {Since $\rho _\alpha $ and $\rho _\beta $ are Hermitian and
  positive semi-definite, we can use their eigendecomposition $\rho _\alpha
  =\DOTSB \sum@ \slimits@ _j a_j \mathinner {|{a_j}\delimiter "526930B
  }\mathinner {\delimiter "426830A {a_j}|}$ and $\rho _\beta =\DOTSB \sum@
  \slimits@ _k b_k \mathinner {|{b_k}\delimiter "526930B }\mathinner
  {\delimiter "426830A {b_k}|}$, with $a_j,b_k \geq 0$, such that $\tr {\rho
  _\alpha \rho _\beta }=\sum _{j,k} a_j b_k | \bracket {a_j}{b_k} |^2\geq
  0$.}\BibitemShut {Stop}%
\bibitem [{Note2()}]{Note2}%
  \BibitemOpen
  \bibinfo {note} {Note that in~\protect \textup {\hbox {\mathsurround \z@
  \protect \normalfont (\ignorespaces \ref {eq:sigma}\unskip \@@italiccorr )}}
  we name the permutations $\sigma $ since (different to $\kappa $ and $\pi $)
  they act on modes rather than particles.}\BibitemShut {Stop}%
\bibitem [{\citenamefont {Baumslag}\ and\ \citenamefont
  {Chandler}(1968)}]{Baumslag-SO-1968}%
  \BibitemOpen
  \bibfield  {author} {\bibinfo {author} {\bibfnamefont {B.}~\bibnamefont
  {Baumslag}}\ and\ \bibinfo {author} {\bibfnamefont {B.}~\bibnamefont
  {Chandler}},\ }\href@noop {} {\emph {\bibinfo {title} {Schaum's outline of
  theory and problems of group theory}}},\ Schaum's outline series\ (\bibinfo
  {publisher} {McGraw-Hill},\ \bibinfo {address} {New York},\ \bibinfo {year}
  {1968})\ \bibinfo {note} {cover title: Theory and problems of group
  theory.}\BibitemShut {Stop}%
\bibitem [{\citenamefont {Ketterer}, \citenamefont {Wyderka},\ and\
  \citenamefont {G{\"u}hne}(2019)}]{Ketterer-CM-2019}%
  \BibitemOpen
  \bibfield  {author} {\bibinfo {author} {\bibfnamefont {A.}~\bibnamefont
  {Ketterer}}, \bibinfo {author} {\bibfnamefont {N.}~\bibnamefont {Wyderka}}, \
  and\ \bibinfo {author} {\bibfnamefont {O.}~\bibnamefont {G{\"u}hne}},\ }\href
  {\doibase 10.1103/PhysRevLett.122.120505} {\bibfield  {journal} {\bibinfo
  {journal} {Phys. Rev. Lett.}\ }\textbf {\bibinfo {volume} {122}},\ \bibinfo
  {pages} {120505} (\bibinfo {year} {2019})}\BibitemShut {NoStop}%
\bibitem [{\citenamefont {Brydges}\ \emph {et~al.}(2019)\citenamefont
  {Brydges}, \citenamefont {Elben}, \citenamefont {Jurcevic}, \citenamefont
  {Vermersch}, \citenamefont {Maier}, \citenamefont {Lanyon}, \citenamefont
  {Zoller}, \citenamefont {Blatt},\ and\ \citenamefont
  {Roos}}]{Brydges-PR-2019}%
  \BibitemOpen
  \bibfield  {author} {\bibinfo {author} {\bibfnamefont {T.}~\bibnamefont
  {Brydges}}, \bibinfo {author} {\bibfnamefont {A.}~\bibnamefont {Elben}},
  \bibinfo {author} {\bibfnamefont {P.}~\bibnamefont {Jurcevic}}, \bibinfo
  {author} {\bibfnamefont {B.}~\bibnamefont {Vermersch}}, \bibinfo {author}
  {\bibfnamefont {C.}~\bibnamefont {Maier}}, \bibinfo {author} {\bibfnamefont
  {B.~P.}\ \bibnamefont {Lanyon}}, \bibinfo {author} {\bibfnamefont
  {P.}~\bibnamefont {Zoller}}, \bibinfo {author} {\bibfnamefont
  {R.}~\bibnamefont {Blatt}}, \ and\ \bibinfo {author} {\bibfnamefont {C.~F.}\
  \bibnamefont {Roos}},\ }\href {\doibase 10.1126/science.aau4963} {\bibfield
  {journal} {\bibinfo  {journal} {Science}\ }\textbf {\bibinfo {volume}
  {364}},\ \bibinfo {pages} {260} (\bibinfo {year} {2019})}\BibitemShut
  {NoStop}%
\bibitem [{\citenamefont {Rademacher}(1922)}]{Rademacher-ES-1922}%
  \BibitemOpen
  \bibfield  {author} {\bibinfo {author} {\bibfnamefont {H.}~\bibnamefont
  {Rademacher}},\ }\href {\doibase 10.1007/BF01458040} {\bibfield  {journal}
  {\bibinfo  {journal} {Math. Ann.}\ }\textbf {\bibinfo {volume} {87}},\
  \bibinfo {pages} {112} (\bibinfo {year} {1922})}\BibitemShut {NoStop}%
\bibitem [{\citenamefont {Walsh}(1923)}]{Walsh-CS-1923}%
  \BibitemOpen
  \bibfield  {author} {\bibinfo {author} {\bibfnamefont {J.~L.}\ \bibnamefont
  {Walsh}},\ }\href {http://www.jstor.org/stable/2387224} {\bibfield  {journal}
  {\bibinfo  {journal} {Am. J. Math.}\ }\textbf {\bibinfo {volume} {45}},\
  \bibinfo {pages} {5} (\bibinfo {year} {1923})}\BibitemShut {NoStop}%
\end{thebibliography}

%

\end{document}